\documentclass[twocolumn]{aastex62}
\received{2021 October 18}
\revised{2021 November 19}
\accepted{2021 November 23}

\shorttitle{Subaru High-$z$ Exploration of Low-Luminosity Quasars (SHELLQs) XVI}
\shortauthors{Matsuoka et al.}

\begin{document}

\title{Subaru High-$z$ Exploration of Low-Luminosity Quasars (SHELLQs). XVI. 69 New Quasars at $5.8 < z < 7.0$}

\correspondingauthor{Yoshiki Matsuoka}
\email{yk.matsuoka@cosmos.ehime-u.ac.jp}

\author{Yoshiki Matsuoka}
\affil{Research Center for Space and Cosmic Evolution, Ehime University, Matsuyama, Ehime 790-8577, Japan.}

\author{Kazushi Iwasawa}
\affil{ICREA and Institut de Ci{\`e}ncies del Cosmos, Universitat de Barcelona, IEEC-UB, Mart{\'i} i Franqu{\`e}s, 1, 08028 Barcelona, Spain.}

\author{Masafusa Onoue}
\affil{Max-Planck-Institut f\"{u}r Astronomie, K\"{o}nigstuhl 17, D-69117, Heidelberg, Germany.}

\author{Takuma Izumi}
\affil{National Astronomical Observatory of Japan, Mitaka, Tokyo 181-8588, Japan.}

\author{Nobunari Kashikawa}
\affil{Department of Astronomy, School of Science, The University of Tokyo, Tokyo 113-0033, Japan.}

\author{Michael A. Strauss}
\affil{Department of Astrophysical Sciences, Princeton University, Peyton Hall, Princeton, NJ 08544, USA.}

\author{Masatoshi Imanishi}
\affil{National Astronomical Observatory of Japan, Mitaka, Tokyo 181-8588, Japan.}
\affil{Department of Astronomical Science, Graduate University for Advanced Studies (SOKENDAI), Mitaka, Tokyo 181-8588, Japan.}

\author{Tohru Nagao}
\affil{Research Center for Space and Cosmic Evolution, Ehime University, Matsuyama, Ehime 790-8577, Japan.}

\author{Masayuki Akiyama}
\affil{Astronomical Institute, Tohoku University, Aoba, Sendai, 980-8578, Japan.}

\author{John D. Silverman}
\affil{Kavli Institute for the Physics and Mathematics of the Universe, WPI, The University of Tokyo, Kashiwa, Chiba 277-8583, Japan.}

\author{Naoko Asami}
\affil{Seisa University, Hakone-machi, Kanagawa, 250-0631, Japan.}

\author{James Bosch}
\affil{Department of Astrophysical Sciences, Princeton University, Peyton Hall, Princeton, NJ 08544, USA.}

\author{Hisanori Furusawa}
\affil{National Astronomical Observatory of Japan, Mitaka, Tokyo 181-8588, Japan.}

\author{Tomotsugu Goto}
\affil{Institute of Astronomy and Department of Physics, National Tsing Hua University, Hsinchu 30013, Taiwan.}

\author{James E. Gunn}
\affil{Department of Astrophysical Sciences, Princeton University, Peyton Hall, Princeton, NJ 08544, USA.}

\author{Yuichi Harikane}
\affil{Institute for Cosmic Ray Research, The University of Tokyo, Kashiwa, Chiba 277-8582, Japan.}

\author{Hiroyuki Ikeda}
\affil{National Institute of Technology, Wakayama College, Gobo, Wakayama 644-0023, Japan.}

\author{Rikako Ishimoto}
\affil{Department of Astronomy, School of Science, The University of Tokyo, Tokyo 113-0033, Japan.}

\author{Toshihiro Kawaguchi}
\affil{Department of Economics, Management and Information Science, Onomichi City University, Onomichi, Hiroshima 722-8506, Japan.}

\author{Nanako Kato}
\affil{Graduate School of Science and Engineering, Ehime University, Matsuyama, Ehime 790-8577, Japan.}

\author{Satoshi Kikuta}
\affil{Center for Computational Sciences, University of Tsukuba, Tsukuba, Ibaraki 305-8577, Japan.}

\author{Kotaro Kohno}
\affil{Institute of Astronomy, The University of Tokyo, Mitaka, Tokyo 181-0015, Japan.}
\affil{Research Center for the Early Universe, University of Tokyo, Tokyo 113-0033, Japan.}

\author{Yutaka Komiyama}
\affil{National Astronomical Observatory of Japan, Mitaka, Tokyo 181-8588, Japan.}
\affil{Department of Astronomical Science, Graduate University for Advanced Studies (SOKENDAI), Mitaka, Tokyo 181-8588, Japan.}


\author{Chien-Hsiu Lee}
\affil{NSF's National Optical-Infrared Astronomy Research Laboratory, 950 North Cherry Avenue, Tucson, AZ 85719, USA.}

\author{Robert H. Lupton}
\affil{Department of Astrophysical Sciences, Princeton University, Peyton Hall, Princeton, NJ 08544, USA.}

\author{Takeo Minezaki}
\affil{Institute of Astronomy, The University of Tokyo, Mitaka, Tokyo 181-0015, Japan.}

\author{Satoshi Miyazaki}
\affil{National Astronomical Observatory of Japan, Mitaka, Tokyo 181-8588, Japan.}
\affil{Department of Astronomical Science, Graduate University for Advanced Studies (SOKENDAI), Mitaka, Tokyo 181-8588, Japan.}


\author{Hitoshi Murayama}
\affil{Kavli Institute for the Physics and Mathematics of the Universe, WPI, The University of Tokyo, Kashiwa, Chiba 277-8583, Japan.}


\author{Atsushi J. Nishizawa}
\affil{Institute for Advanced Research, Nagoya University, Furo-cho, Chikusa-ku, Nagoya 464-8602, Japan.}


\author{Masamune Oguri}
\affil{Department of Physics, Graduate School of Science, The University of Tokyo, Bunkyo, Tokyo 113-0033, Japan}
\affil{Kavli Institute for the Physics and Mathematics of the Universe, WPI, The University of Tokyo, Kashiwa, Chiba 277-8583, Japan.}
\affil{Research Center for the Early Universe, University of Tokyo, Tokyo 113-0033, Japan.}

\author{Yoshiaki Ono}
\affil{Institute for Cosmic Ray Research, The University of Tokyo, Kashiwa, Chiba 277-8582, Japan}

\author{Masami Ouchi}
\affil{Institute for Cosmic Ray Research, The University of Tokyo, Kashiwa, Chiba 277-8582, Japan}
\affil{Kavli Institute for the Physics and Mathematics of the Universe, WPI, The University of Tokyo, Kashiwa, Chiba 277-8583, Japan.}

\author{Paul A. Price}
\affil{Department of Astrophysical Sciences, Princeton University, Peyton Hall, Princeton, NJ 08544, USA.}

\author{Hiroaki Sameshima}
\affil{Institute of Astronomy, The University of Tokyo, Mitaka, Tokyo 181-0015, Japan.}



\author{Naoshi Sugiyama}
\affil{Kavli Institute for the Physics and Mathematics of the Universe, WPI, The University of Tokyo, Kashiwa, Chiba 277-8583, Japan.}
\affil{Graduate School of Science, Nagoya University, Furo-cho, Chikusa-ku, Nagoya 464-8602, Japan.}

\author{Philip J. Tait}
\affil{Subaru Telescope, Hilo, HI 96720, USA.}

\author{Masahiro Takada}
\affil{Kavli Institute for the Physics and Mathematics of the Universe, WPI, The University of Tokyo, Kashiwa, Chiba 277-8583, Japan.}

\author{Ayumi Takahashi}
\affil{Graduate School of Science and Engineering, Ehime University, Matsuyama, Ehime 790-8577, Japan.}

\author{Tadafumi Takata}
\affil{National Astronomical Observatory of Japan, Mitaka, Tokyo 181-8588, Japan.}
\affil{Department of Astronomical Science, Graduate University for Advanced Studies (SOKENDAI), Mitaka, Tokyo 181-8588, Japan.}

\author{Masayuki Tanaka}
\affil{National Astronomical Observatory of Japan, Mitaka, Tokyo 181-8588, Japan.}
\affil{Department of Astronomical Science, Graduate University for Advanced Studies (SOKENDAI), Mitaka, Tokyo 181-8588, Japan.}


\author{Yoshiki Toba}
\affil{Department of Astronomy, Kyoto University, Sakyo-ku, Kyoto, Kyoto 606-8502, Japan.}

\author{Yousuke Utsumi}
\affil{Kavli Institute for Particle Astrophysics and Cosmology, Stanford University, CA 94025, USA.}

\author{Shiang-Yu Wang}
\affil{Institute of Astronomy and Astrophysics, Academia Sinica, Taipei, 10617, Taiwan.}

\author{Takuji Yamashita}
\affil{National Astronomical Observatory of Japan, Mitaka, Tokyo 181-8588, Japan.}



\begin{abstract}

We present the spectroscopic discovery of 69 quasars at $5.8 < z < 7.0$, drawn from the Hyper Suprime-Cam (HSC) Subaru Strategic Program (SSP) imaging survey data.
This is the 16th publication from the Subaru High-$z$ Exploration of Low-Luminosity Quasars (SHELLQs) project, and completes identification of all but the faintest
candidates (i.e., $i$-band dropouts with $z_{\rm AB} < 24$ and $y$-band detections, and $z$-band dropouts with $y_{\rm AB} < 24$) 
with Bayesian quasar probability $P_{\rm Q}^{\rm B} > 0.1$ in the HSC-SSP third public data release (PDR3).
The sample reported here also includes three quasars with $P_{\rm Q}^{\rm B} < 0.1$ at $z \sim 6.6$, which we selected in an effort to completely cover 
the reddest point sources with simple color cuts.
The number of high-$z$ quasars discovered in SHELLQs has now grown to 162, including 23 type-II quasar candidates.
This paper also presents identification of seven galaxies at $5.6 < z < 6.7$, an [\ion{O}{3}] emitter at $z = 0.954$, and 31 Galactic cool stars and brown dwarfs.
High-$z$ quasars and galaxies comprise 75 \% 
and 16 \% respectively of all the spectroscopic SHELLQs objects that pass
our latest selection algorithm with the PDR3 photometry.
That is, a total of 91 \% of the objects lie at $z > 5.6$. 
This demonstrates that the algorithm has very high efficiency, even though we are probing an unprecedentedly low-luminosity population down to $M_{1450} \sim -21$ mag.

\end{abstract}


\section{Introduction} \label{sec:intro}

The astronomical community is making great strides toward charting and understanding quasars at the epoch of cosmic reionization, which is thought to have taken place during the interval $6 < z < 10$ \citep{planck20}.
Quasars at $5.7 < z \le 7.0$ have been discovered from optical wide-field ($>$100-deg$^2$ class) multi-band surveys, such as the Sloan Digital Sky Survey 
\citep[SDSS;][]{fan00,fan01,fan03,fan04,fan06, jiang08,jiang09,jiang15},
the Canada-France-Hawaii Telescope Legacy Survey \citep[CFHTLS;][]{willott05,willott07,willott09,willott10a,willott10b}, 
the Panoramic Survey Telescope And Rapid Response System 1 survey \citep[Pan-STARRS1;][]{banados14,banados16,venemans15a,mazzucchelli17}, 
the Dark Energy Survey \citep[DES;][]{reed15, reed17, reed19, yang19}, and the Dark Energy Spectroscopic Instrument Legacy Imaging Surveys
\citep[DELS;][]{wang17, wang18, wang19}.
Near-infrared (IR) surveys are paving the way to probe more distant objects, and indeed quasars at $z > 6.5$ have been detected with, e.g., 
the United Kingdom Infrared Telescope (UKIRT) Infrared Deep Sky Survey \citep{mortlock11}, 
the Visible and Infrared Survey Telescope for Astronomy (VISTA) Kilo-degree Infrared Galaxy Survey \citep[VIKING;][]{venemans13}, 
and the UKIRT and VISTA Hemisphere Surveys,
with three objects at $z = 7.5 - 7.6$ marking the highest-redshift quasars currently known \citep{banados18, yang20, wang21}.
We will soon reach deeper into the epoch of reionization with the advent of {\it Euclid}  
and the {\it Roman Space Telescope}.
These two space missions will provide unprecedentedly wide-and-deep maps of the near-IR sky, 
and are expected to identify quasars up to 
$z \sim 10$ \citep[e.g.,][]{euclid19}.
High-$z$ quasars thus discovered have been probing, and will further probe, 
the formation of the first supermassive black holes (SMBHs) and their host galaxies, the history and sources of reionization, and
other key issues in the early universe.

We have been carrying out a high-$z$ quasar survey complementary to the existing ones for the past several years, going much deeper in relatively small areas
of the sky.
The project (``Subaru High-$z$ Exploration of Low-Luminosity Quasars"; SHELLQs) exploits the exquisite imaging data produced by the Hyper Suprime-Cam \citep[HSC;][]{miyazaki18} 
Subaru Strategic Program (SSP) survey.
Three layers named Wide, Deep, and UltraDeep constitute the HSC-SSP survey, 
covering (1400, 26, 3.5) deg$^2$ down to 5$\sigma$ limiting magnitudes of $i_{\rm AB}$ = (25.9, 26.8, 27.4) for point sources, 
respectively \citep{aihara17_survey}.
Thus far we have reported spectroscopic identification of 93 low-luminosity quasars at $5.7 < z < 7.1$, including 18 type-II quasar candidates  with very luminous and narrow Ly$\alpha$ 
emission, in a series of SHELLQs publications \citep{p1, p2, p4, p10, p7}.
Follow-up near-IR spectroscopy has revealed that the quasars have a variety of accretion modes, from sub-Eddington accretion onto
massive SMBHs to (super-)Eddington accretion onto less massive ones \citep[][M. Onoue et al., in prep.]{onoue19}.
The host galaxies probed with ALMA are also diverse, sometimes accompanied by active star formation and powerful extended outflows
\citep{izumi21a, izumi21b}, while their dynamical masses are more or less consistent with those inferred from 
the local mass relation between SMBHs and the host bulges
\citep{izumi18, izumi19}.

This is the 16th publication from the SHELLQs project, presenting the spectroscopic discovery of 69 new high-$z$ quasars accumulated over the past two years.
The basics of candidate selection and spectroscopic observations are described in \S \ref{sec:obs}, followed by the results and discussion 
in \S \ref{sec:results}.
We use point-spread-function (PSF) magnitudes ($m_{\rm AB}$) and associated errors ($\sigma_{\rm m})$ presented in the AB system \citep{oke83}
unless otherwise noted, 
corrected for Galactic extinction \citep{schlegel98}.
We refer to $z$-band magnitudes with the AB subscript (``$z_{\rm AB}$"), while redshift $z$ appears without a subscript.
The cosmological parameters are assumed to be $H_0$ = 70 km s$^{-1}$ Mpc$^{-1}$, $\Omega_{\rm M}$ = 0.3, and $\Omega_{\rm \Lambda}$ = 0.7.

\section{Candidate selection and spectroscopy} \label{sec:obs}

Our candidate selection strategy remains mostly unchanged since the beginning of the project, and is detailed in our previous papers.
Here, we itemize the essential steps, with an emphasis on recent updates.
\begin{enumerate}

\item We first select point sources meeting the criteria\\
($z_{\rm AB} < 24.5$ \& $\sigma_z < 0.155$ \& $i_{\rm AB} - z_{\rm AB} > 1.5$) OR\\
($y_{\rm AB} < 25.0$ \& $\sigma_y < 0.217$ \& $z_{\rm AB} - y_{\rm AB} > 0.8$)\\ 
from the HSC-SSP database.
The candidates selected with the first/second set of conditions are referred to as $i$-/$z$-dropouts in what follows.
Our definition of ``point sources" uses 
the ratio between the adaptive moment \citep{hirata03} of a given source ($\mu$; averaged over the two image dimensions) and 
that of the PSF model ($\mu_{\rm PSF}$).
For $i$-dropouts, we use the cut $0.7 < \mu/\mu_{\rm PSF} < 1.2$ measured in the $z$-band.
This criterion 
removes spectroscopic high-$z$ galaxies more efficiently 
than does the condition we previously used 
(i.e., requiring that the difference between the PSF and CModel magnitudes be less than 0.15),
while retaining $>$90 \% of the high-$z$ broad-line quasars we identified \citep[see Figure 8 of][]{p10}.\footnote{
The criterion is determined as a compromise between the requirements to have high completeness and low contamination rates.
A higher value of the maximum allowed $\mu/\mu_{\rm PSF}$ selects quasars with larger contribution of the host galaxies 
\citep[e.g.,][]{boutsia21, bowler21}
as well as more galaxies without quasars \citep[e.g.,][]{ono18}.
}
A looser cut, $0.6 < \mu/\mu_{\rm PSF} < 3.0$ measured in the $y$-band, is used for $z$-dropouts, to be more inclusive.
We exclude from the selection those sources with $g$- or $r$-band detections or any critical quality issues in the photometry, such as those caused by saturation, 
cosmic rays, bad pixels, and nearby bright stars.
In addition, we eliminate the bluest candidates with $i_{\rm AB} - z_{\rm AB} < 2.5$ and $z_{\rm AB} - y_{\rm AB} < -1.0$, as we have found that such sources 
are dominated by [\ion{O}{3}] line emitters at $z \sim 0.8$ (see below).

\item The list of selected sources is matched to public near-IR survey catalogs.
All the objects presented in this paper were found from the HSC-SSP Wide layer, where we use $Y$-, $J$-, $H$-, and $K$-band magnitudes 
measured with the UKIRT Wide-Field Camera \citep[WFCAM;][]{casali07} or the VISTA Infrared Camera \citep[VIRCAM;][]{dalton06}.
The data were obtained as a part of the UKIDSS Large Area Survey (available in the data release [DR] 11PLUS), the UKIDSS Hemisphere Survey (DR1), 
the VIKING (DR5), or the VISTA Hemisphere Survey (DR6).
The entire HSC-SSP Wide layer is covered in at least one band of the above NIR surveys.

\item We calculate a Bayesian probability ($P_{\rm Q}^{\rm B}$) for each candidate being a high-$z$ quasar, following the recipe provided by \citet{mortlock12}, 
using the matched HSC $i$-, $z$-, $y$-band and near-IR magnitudes.
The calculation considers flux upper limits in the HSC bands, while near-IR magnitudes are used only when a source is detected in the given band.
Our algorithm includes models of high-$z$ quasars and Galactic cool stars and brown dwarfs, taking into account 
their spectral energy distributions (SEDs) and surface densities as a function of apparent magnitude and Galactic coordinates.
We select all the candidates with $P_{\rm Q}^{\rm B} > 0.1$, and also keep a fraction of the remaining sources 
(the ``low-$P_{\rm Q}^{\rm B}$ sample" hereafter; see below), for the subsequent selection.

\item All the pre-stacked and stacked images of the candidates are retrieved from the HSC-SSP database, and screened by an automatic
algorithm based on Source Extractor \citep{bertin96} and by visual inspection. 
This step removes numerous false detections missed by the catalog quality flags, such as cosmic rays and detector artifacts,
as well as variable sources.

\item The candidates are fed into follow-up spectroscopy programs at the 8.2-m Subaru Telescope and the 10.4-m Gran Telescopio Canarias (GTC).
We use the Faint Object Camera and Spectrograph \citep[FOCAS;][]{kashikawa02} on Subaru to observe the wavelength range 0.75 -- 1.0 $\mu$m with
the VPH900 grism and 1\arcsec.0 slits, giving a spectral resolution of $R \sim 1200$ (Program ID: S18B-011I). 
Similarly, the Optical System for Imaging and low-intermediate-Resolution Integrated Spectroscopy \citep[OSIRIS;][]{cepa00} on GTC 
is used with the R2500I grism and 1\arcsec.0 longslit, yielding spectra over 0.75 -- 1.0 $\mu$m with $R \sim 1500$ 
(Program IDs: GTC5-20A, GTC10-20B, GTC42-21B). 
The objects presented in this paper were observed mostly in gray nights under both photometric and non-photometric sky conditions, 
with a typical seeing of 0\arcsec.6 -- 1\arcsec.2.

\end{enumerate}

We have so far completed spectroscopy of all but the faintest candidates drawn from $\sim$1200 deg$^2$ of the HSC-SSP public data release 3 
\citep[PDR3;][]{aihara21}.
The remaining candidates are either 
(i) $i$-dropouts with $z_{\rm AB} > 24.0$ or without $y$-band detections, or (ii) $i$-dropouts with $z_{\rm AB} > 23.5$ and $i_{\rm AB} - z_{\rm AB} < 2.0$.
All the $z$-dropout candidates have been observed;
while the formal magnitude cut is $y_{\rm AB} = 25.0$ (Step 1), the actual limiting magnitude depends mostly on the depth achieved in the HSC imaging,
which is $y_{\rm AB} \sim 24.4$ for 5$\sigma$ detection of point sources in the Wide layer \citep{aihara21}.
This paper presents spectroscopic identification of 108 candidates, 
observed
over the past two years since our previous discovery paper \citep{p10}.

\section{Results and Discussion \label{sec:results}}

\begin{figure*}
\epsscale{0.97}
\plotone{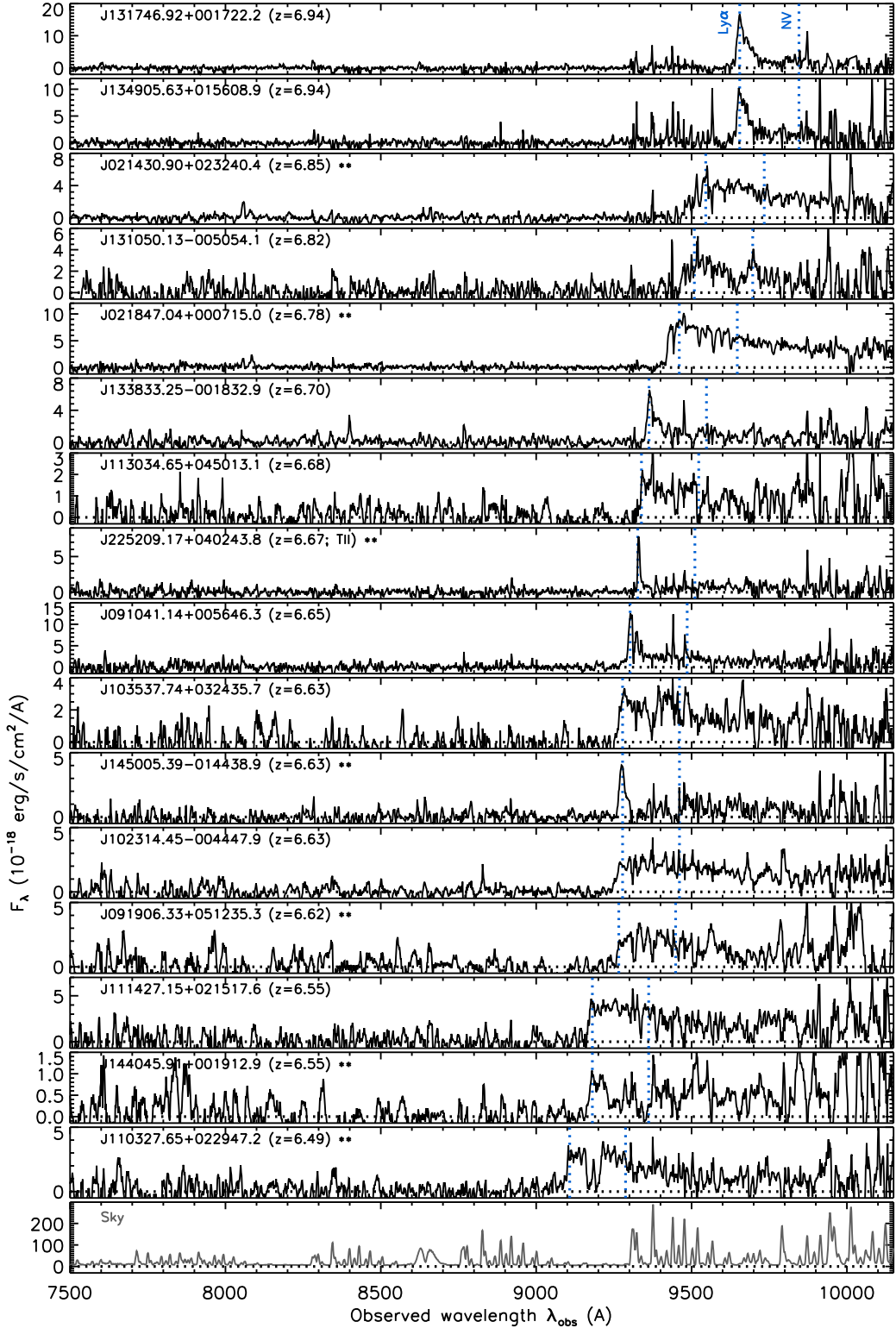}
\caption{Discovery spectra of the first set of 16 quasars, displayed in decreasing order of redshift.
The object name and the estimated redshift (with uncertainty $\Delta z \sim 0.01 - 0.1$; see text) are indicated at the top left corner of each panel,
with the two asterisks indicating that the object is noted in the text.
The five type-II quasar candidates are marked with ``TII" adjacent to the redshifts.
The blue dotted lines mark the expected positions of the Ly$\alpha$ and \ion{N}{5} $\lambda$1240 emission lines, given the redshifts.
The spectra were smoothed using inverse-variance weighted means over 3 -- 13 pixels (depending on the signal-to-noise ratio [S/N]), for display purposes.
The bottom panel displays a sky spectrum, as a guide to the expected noise.
\label{fig:spectra1}}
\end{figure*}

\begin{figure*}
\epsscale{1.0}
\plotone{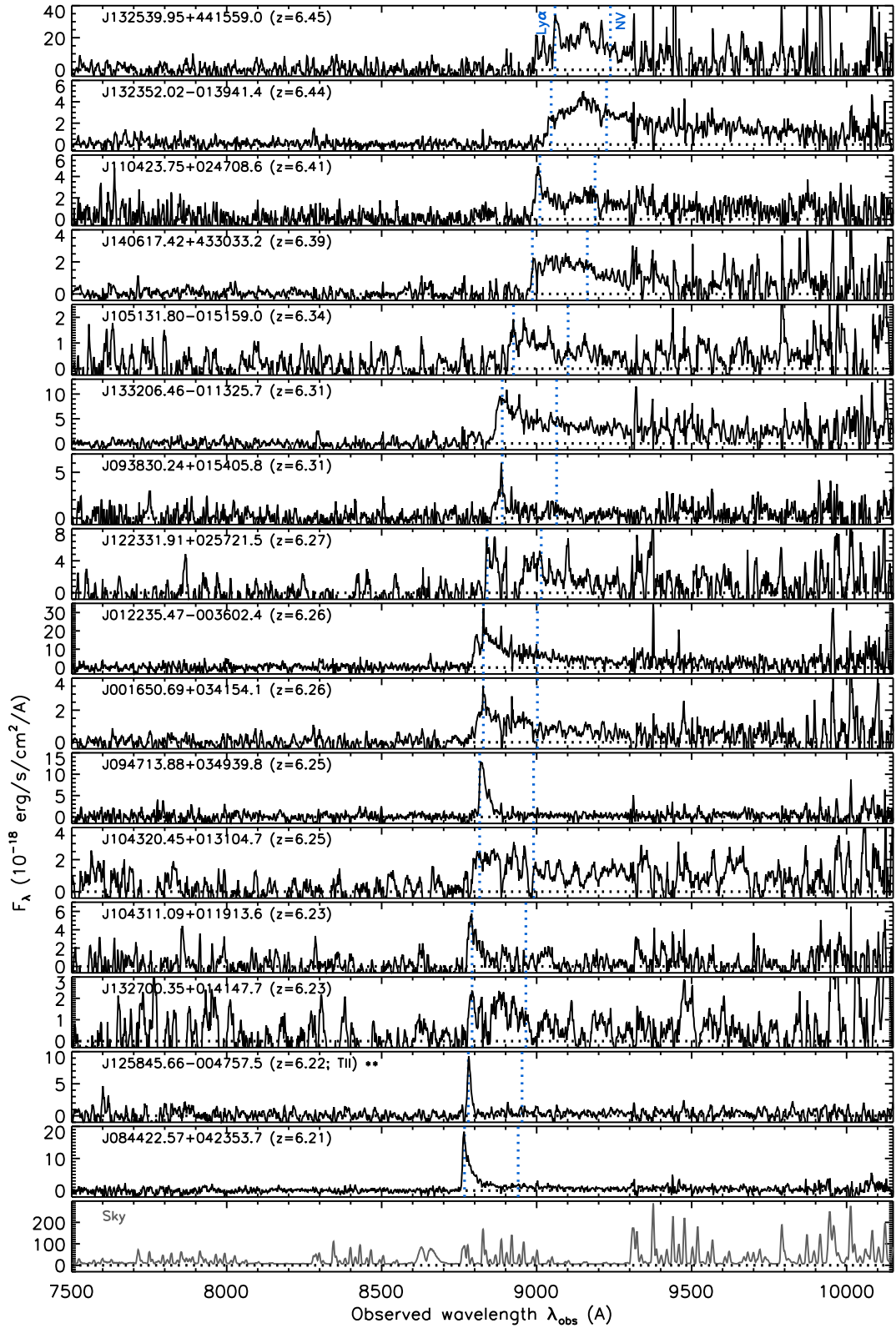}
\caption{Same as Figure \ref{fig:spectra1}, but for the second set of 16 quasars.
\label{fig:spectra2}}
\end{figure*}

\begin{figure*}
\epsscale{1.0}
\plotone{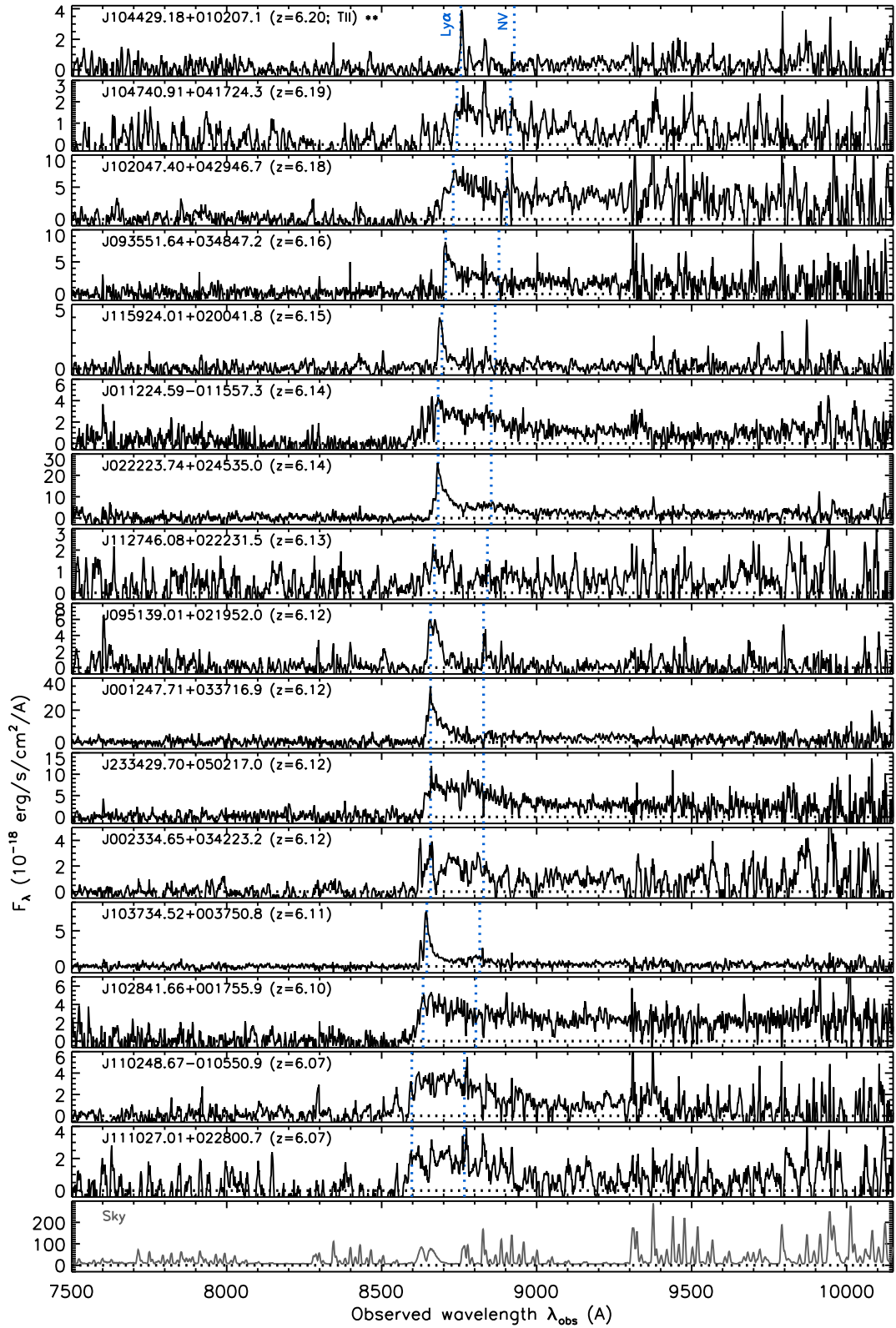}
\caption{Same as Figure \ref{fig:spectra1}, but for the third set of 16 quasars.
\label{fig:spectra3}}
\end{figure*}

\begin{figure*}
\epsscale{1.0}
\plotone{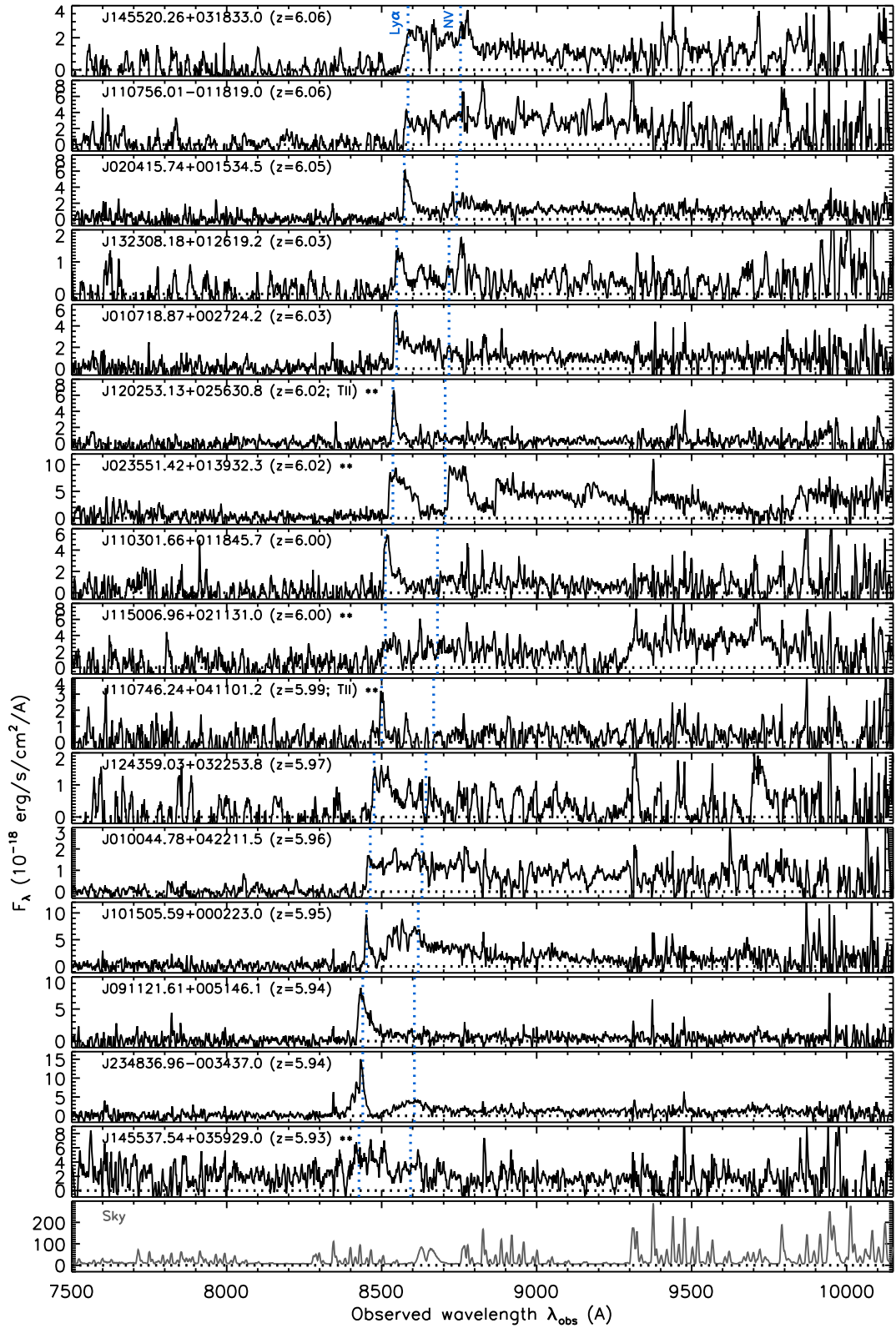}
\caption{Same as Figure \ref{fig:spectra1}, but for the fourth set of 16 quasars.
\label{fig:spectra4}}
\end{figure*}

\begin{figure*}
\epsscale{1.0}
\plotone{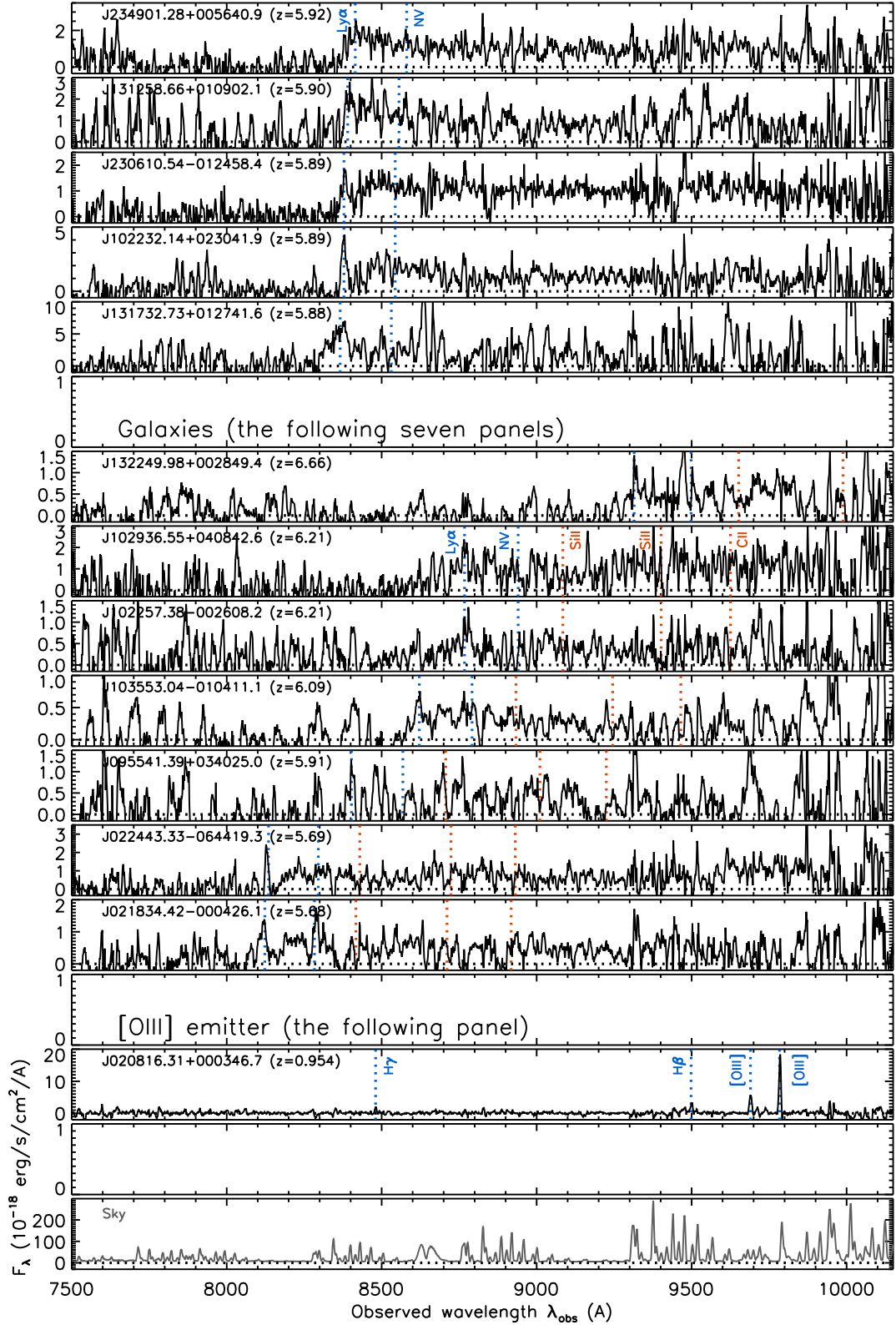}
\caption{Same as Figure \ref{fig:spectra1}, but for the last set of quasars in the top five panels.
The following panels present seven galaxies and an [\ion{O}{3}] emitter,
with the expected positions of \ion{Si}{2} $\lambda$1260, \ion{Si}{2} $\lambda$1304, and \ion{C}{2} $\lambda$1335 (for the galaxies) 
and of H$\gamma$, H$\beta$, and [\ion{O}{3}] $\lambda$4959, $\lambda$5007 lines (for the [\ion{O}{3}] emitter) marked by the dotted lines.
\label{fig:spectra5}}
\end{figure*}

\begin{figure*}
\epsscale{1.0}
\plotone{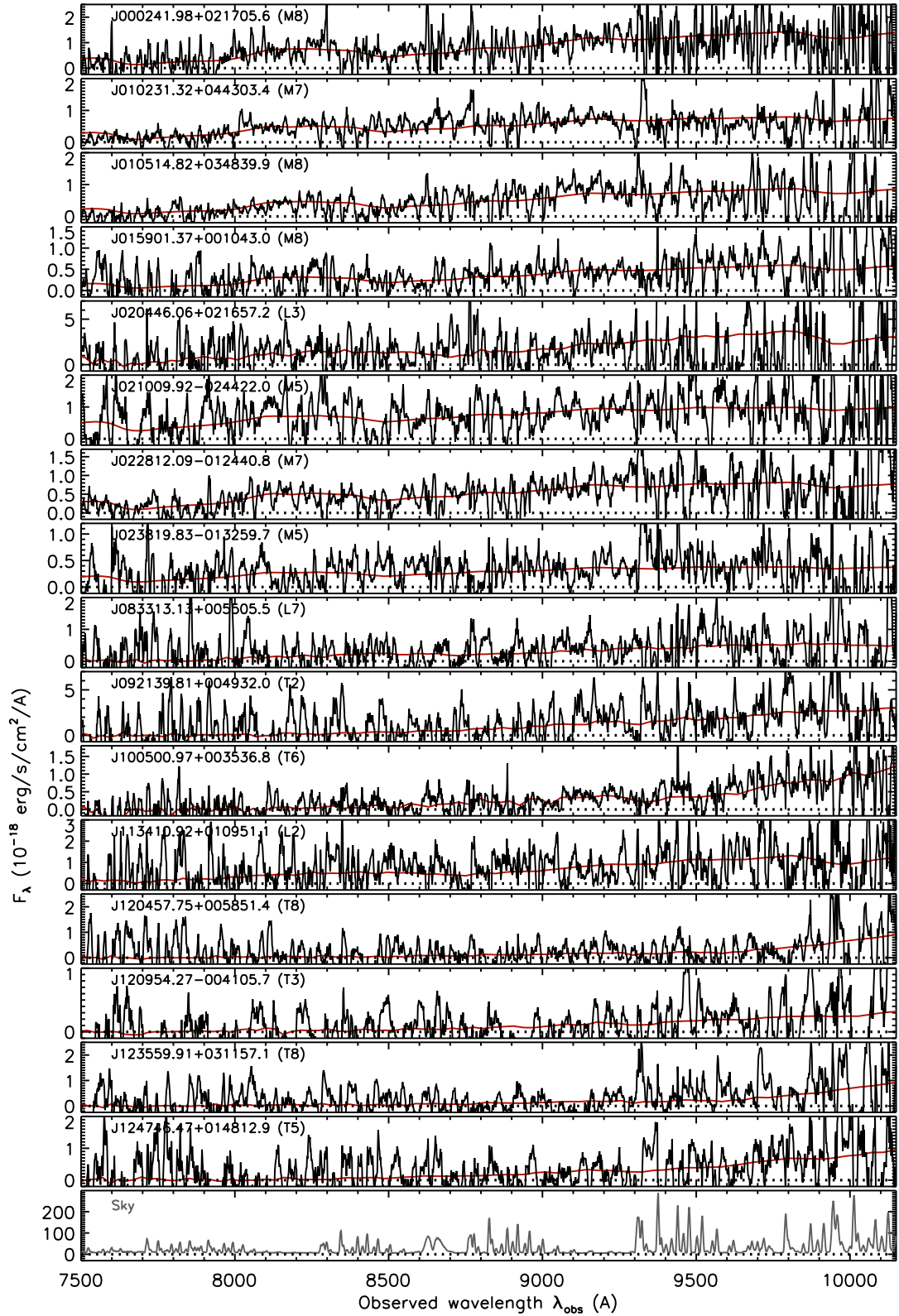}
\caption{Same as Figure \ref{fig:spectra1}, but for the first set of 16 Galactic cool stars and brown dwarfs, ordered by the right ascension. 
The red lines represent the best-fit templates, whose spectral types are indicated at the top left corner of the individual panels.
\label{fig:spectra6}}
\end{figure*}

\begin{figure*}
\epsscale{1.0}
\plotone{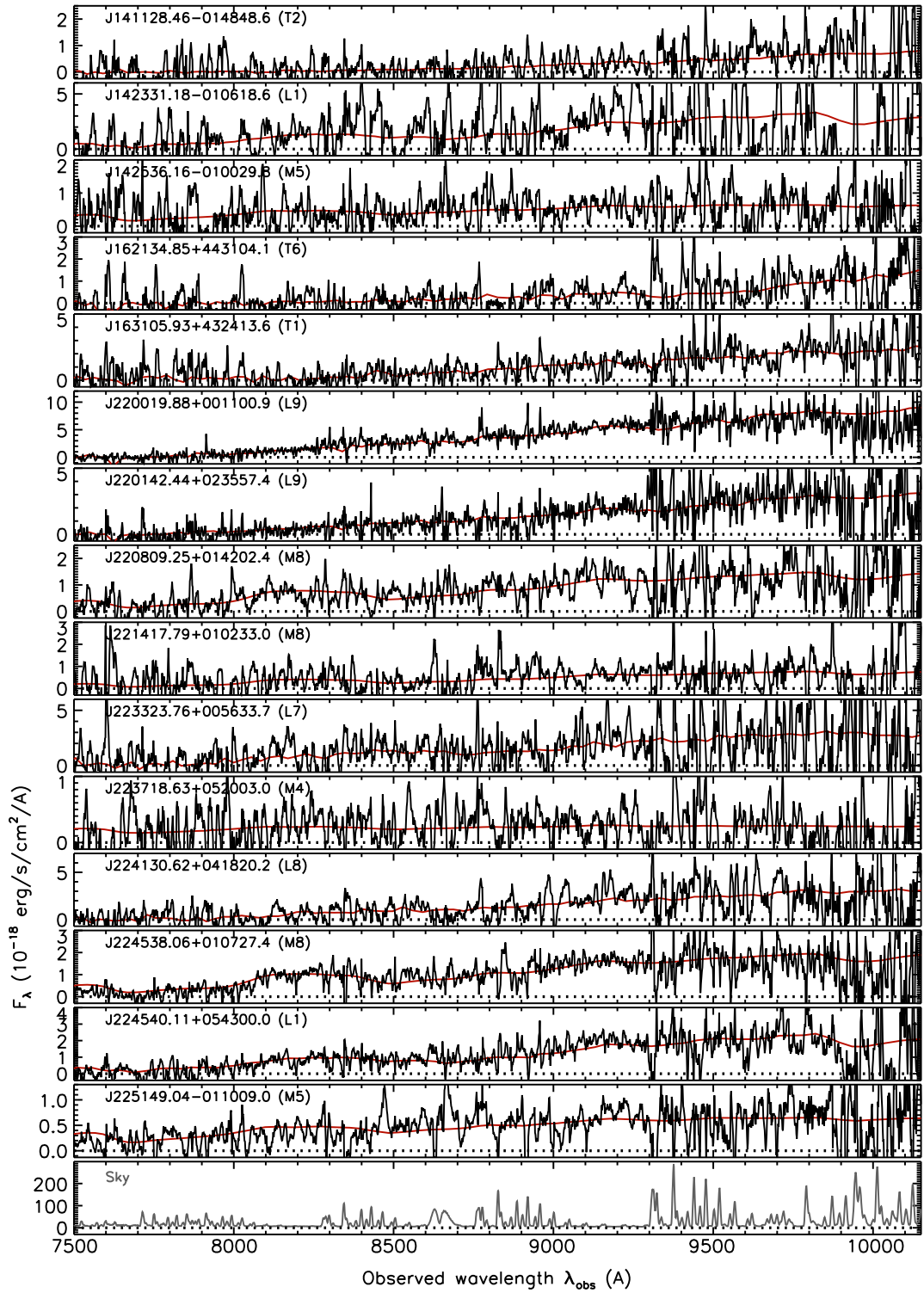}
\caption{Same as Figure \ref{fig:spectra6}, but for the last set of 15 Galactic dwarfs. 
\label{fig:spectra7}}
\end{figure*}

Table \ref{tab:obsjournal} is the journal of discovery spectroscopy, including the coordinates and photometric information of the observed candidates.
The objects detected in the 
near-IR bands are listed in Table \ref{tab:nir_photometry}.
The 108 candidates include 
69 quasars at $5.8 < z < 7.0$, seven galaxies at $5.6 < z < 6.7$, an [\ion{O}{3}] emitter at $z = 0.954$, and 31 cool stars and brown dwarfs in the Milky Way.
Figures \ref{fig:spectra1} -- \ref{fig:spectra7} present their reduced spectra, in the same order as in Table \ref{tab:obsjournal}.
The spectra have been scaled in flux to match the HSC magnitudes in the $z$/$y$ bands for $i$-/$z$-dropouts.
We found clear trace of signals in both the 2d and 1d spectra of all the presented objects.

The classification and measurements of the spectra are performed in a way consistent with our previous papers.
The 69 objects in Figures \ref{fig:spectra1} -- \ref{fig:spectra5} (top panels) are classified as high-$z$ quasars based on the broad Ly$\alpha$ line, 
blue rest-frame ultraviolet continuum, and/or sharp continuum break just blueward of Ly$\alpha$.
Seven high-$z$ objects lacking AGN signatures are classified as galaxies (Figure \ref{fig:spectra5} middle panels).
The 69 quasars include five type-II candidates ($J225209.17+040243.8$, $J125845.66-004757.5$, $J104429.18+010207.1$, $J120253.13+025630.8$, and $J110746.24+041101.2$) 
with very luminous ($>10^{43}$ erg s$^{-1}$) and narrow (full width at half maximum [FWHM] $<$ 500 km s$^{-1}$) Ly$\alpha$ emission. 
Such Ly$\alpha$ features are often associated with active galactic nuclei (AGNs) in the lower-$z$ Universe \citep[e.g.,][]{alexandroff13, konno16, sobral18, spinoso20}.
Indeed, our deep Keck/MOSFIRE spectroscopy of a similar object from SHELLQs has revealed very strong \ion{C}{4} $\lambda$1549 doublet lines, 
demonstrating that it is an AGN \citep{onoue21}.
Past surveys of high-$z$ Ly$\alpha$ emitters (LAEs) also found similar objects at the bright end ($>10^{43}$ erg s$^{-1}$) of the Ly$\alpha$ luminosity function, 
with a significant contribution expected from AGNs \citep[e.g.,][]{santos16, konno18}.
However, it is currently hard to make a clear distinction between extreme star formation, AGN, and other possibilities powering those most luminous LAEs; 
future observations in other wavelengths, in particular X-rays for AGNs, will be key to making robust classifications.

\begin{figure}
\epsscale{1.18}
\plotone{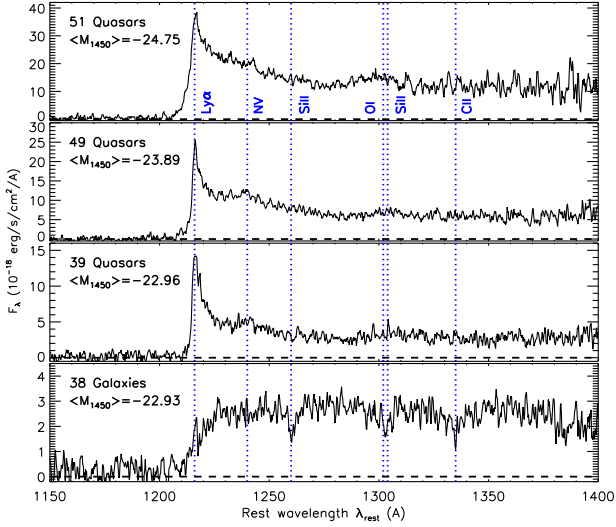}
\caption{Median-stacked spectra of the high-$z$ objects discovered by SHELLQs.
The top three panels present those created from 51 quasars with $M_{1450} < -24.2$, 49 quasars with $-23.5 < M_{1450} < -24.2$, 
and 39 quasars with $M_{1450} > -23.5$, while the bottom panel presents that created from 38 galaxies.
The median $M_{1450}$ of the stacked objects is reported in each panel.
\label{fig:stack}}
\end{figure}

The distinction between the faint (type-I) quasars and galaxies is also sometimes ambiguous, partly due to the limited data quality.
Figure \ref{fig:stack} presents the composite spectra of the high-$z$ quasars in three bins of absolute magnitudes ($M_{1450} < -24.2$, $-24.2 < M_{1450} < -23.5$, 
and $-23.5 < M_{1450}$; type-II candidates were excluded) and of the high-$z$ galaxies, discovered in SHELLQs so far.
These spectra were generated by moving the individual spectra to the rest frame, normalizing to the median $M_{1450}$ of each group of objects being stacked, 
and then median-stacking. 
The 39 quasars in the lowest luminosity bin and the 38 galaxies have similar distributions and median values of $M_{1450}$, but the composite spectra are
strikingly different; the quasar composite shows clear broad emission lines of Ly$\alpha$ and \ion{N}{5} $\lambda$1240, while the galaxy composite lacks such emission features
and instead has interstellar absorption 
lines of \ion{Si}{2} $\lambda$1260, \ion{Si}{2} $\lambda$1304, and \ion{C}{2} $\lambda$1335.
This demonstrates that our spectral separation between quasars and galaxies is robust as a whole, while there may be minor cases of misclassification for individual objects. 

\begin{figure}
\epsscale{1.2}
\plotone{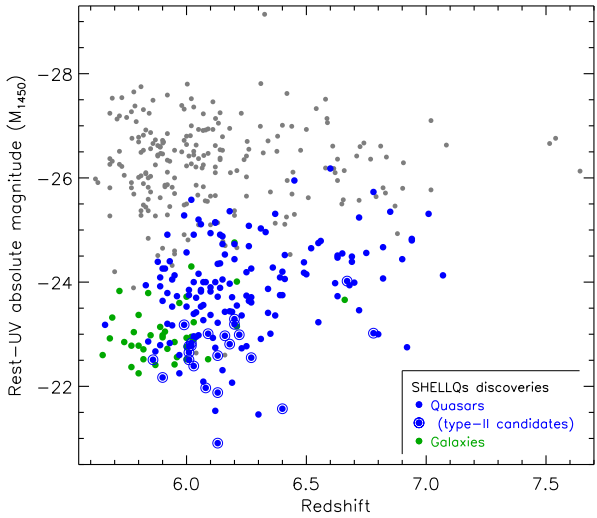}
\caption{Redshifts and the rest-frame ultraviolet absolute magnitudes ($M_{1450}$) of all the high-$z$ objects reported in the past and present
SHELLQs papers, including quasars ({\it blue dots}), type-II quasar candidates ({\it blue dots enclosed by larger circles}), and galaxies ({\it green dots}).
We also plot all the other high-$z$ quasars published to date ({\it gray dots}), compiled mostly from the papers referenced in \S \ref{sec:intro}.
\label{fig:zM1450}}
\end{figure}

We measured the redshift of each quasar or galaxy via the observed wavelength of the Ly$\alpha$ line or of the onset of the IGM \ion{H}{1} absorption.
This procedure is not always easy, due to the \ion{H}{1} damping-wing absorption, the ambiguity in determining the onset of the \citet[][GP]{gunn65} trough, and 
low signal-to-noise (S/N) spectra in some cases.
The uncertainty of the present redshift measurements is thus relatively large, from $\Delta z \sim 0.01$ (when Ly$\alpha$ has a clear peak) to $\sim 0.1$ 
(when Ly$\alpha$ emission is not visible).
The absolute magnitude ($M_{1450}$) of each object was measured by extrapolating the observed luminosity in a continuum window, selected at wavelengths relatively free of
strong sky emission lines, by assuming a power-law continuum model ($F_{\lambda} \propto \lambda^{\alpha}$) with
$\alpha = -1.5$ for quasars \citep[e.g.,][]{vandenberk01} or $\alpha = -2.0$ for galaxies \citep[e.g.,][]{stanway05}.
The Ly$\alpha$ ($+$ \ion{N}{5} $\lambda$1240) properties were measured with the continuum flux estimated either on the red side of the line (for objects 
with relatively narrow Ly$\alpha$) or with the above power-law continuum model (for the remaining objects).
Table \ref{tab:spectroscopy} summarizes the results of the spectral measurements, and Figure \ref{fig:zM1450} presents the distribution of redshifts and absolute magnitudes
of all the SHELLQs objects at $z > 5.6$, including those reported in our previous papers.

\begin{figure}
\epsscale{1.22}
\plotone{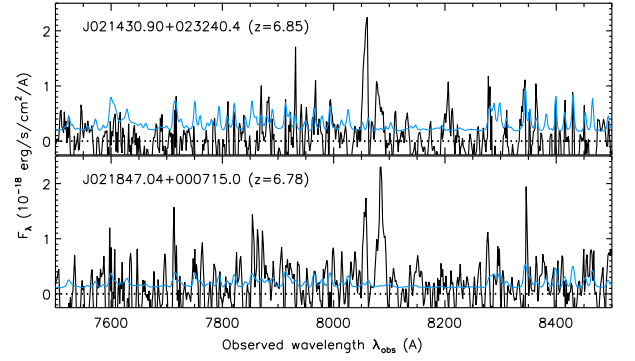}
\caption{Examples of the transmitted fluxes in the GP trough, found along the sightlines toward J021430.90+023240.4 (top) and J021847.04+000715.0 (bottom).
The black and cyan lines represent the observed fluxes and associated 1$\sigma$ errors, respectively.
\label{fig:IGMtransmission}}
\end{figure}

Here we briefly note on a few quasars with unusual spectral features.  
$J115006.96+021131.0$ ($z = 6.00$) has a second continuum break at $\lambda_{\rm obs} \simeq 9300$ \AA.
This break is also clear in the 2d spectra, and the HSC color of this object ($z_{\rm AB} - y_{\rm AB} = 0.64$) is considerably redder than expected for a typical quasar at the same redshift
(see Figure \ref{fig:twocolor} below).
No particular feature is known at the corresponding rest-frame wavelength ($\lambda_{\rm rest} \sim 1330$ \AA) in a typical quasar spectrum
\citep[e.g.,][]{vandenberk01},
and thus the origin of this break is unclear.
The HSC images show no evidence of an overlapping source; the quasar is a clear point source without significant displacement observed
between the $i$-, $z$-, and $y$-band image centroids.
Nonetheless, if the second break corresponds to the Ly$\alpha$ wavelength of a background source, its redshift would be $z \sim 6.6$.
Alternatively, this feature might be caused by a broad absorption line (BAL) of \ion{Si}{2} $\lambda$1398 with very high blueshift velocities, $>$15,000 km s$^{-1}$.
$J023551.42+013932.3$ ($z = 6.02$) exhibits unambiguous features of BALs; weaker absorption features may be present in other objects as well 
(e.g., $J110327.65+022947.2$ at $z = 6.49$), but higher-quality data are necessary for the confirmation of their presence and nature, e.g., whether they are intrinsic
to the quasars or are produced by foreground metal absorbers.
$J145537.54+035929.0$ ($z = 5.93$) has significant excess flux at $\lambda_{\rm obs} < 8400$ \AA~ in the GP trough.
This is surprising, since the HSC colors ($i_{\rm AB} - z_{\rm AB} = 2.1$ and $z_{\rm AB} - y_{\rm AB} = -0.4$) are perfectly consistent
with being a quasar at $z = 5.9$.
Indeed the color estimated from the spectrum, $i_{\rm AB} - z_{\rm AB} < 0.6$,\footnote{
An upper limit is reported here since the spectrum only partially covers the HSC $i$-band. } 
conflicts with the HSC measurement.
This object is either (i) a high-$z$ quasar with an overlapping foreground transient source that appeared after the HSC imaging and affected only the spectroscopic observations,
or (ii) a low-$z$ variable source 
(however, we are not aware of any similar spectra in the literature) 
which was relatively faint/bright at the time of the HSC $i$-/$z$-band imaging, making 
it appear to be an $i$-dropout in broad-band photometry.
We tentatively keep this object in the quasar category for now, and will revisit its nature with future observations.
Finally, while most of our low-luminosity quasars have no observable signal in the GP trough, the brightest ones sometimes exhibit spikes of transmitted flux.
For example, $J021430.90+023240.4$ ($z = 6.85$) and $J021847.04+000715.0$ ($z = 6.78$) have significant positive signals at
$\lambda_{\rm obs} = 8050 - 8100$ \AA\ as displayed in Figure \ref{fig:IGMtransmission}; these signals are also apparent in the 2d spectra.
The large numbers of low-luminosity quasars will offer a unique probe of the small-scale structure of reionization, once higher-quality
spectra are obtained with future deep observations.

Figure \ref{fig:twocolor} displays the two-color diagram of all 312 objects with spectroscopic identification reported in the past and present SHELLQs papers.\footnote{
162 high-$z$ quasars (including 23 type-II candidates), 38 high-$z$ galaxies, 17 [\ion{O}{3}] emitters at $z = 0.8 - 1.0$, and 95 Galactic cool stars and brown dwarfs.
The brown dwarf $J161042.47+554203.4$ presented in \citet{p10} is not found in the PDR3 source catalog for unknown reasons,
and thus is not plotted.}
The quasars populate the lower-right and upper portions of this diagram
and are almost absent in between ($1 < z_{\rm AB} - y_{\rm AB} < 2$), where the quasar model track intersects the stellar and brown-dwarf sequence;
sources with such moderate colors 
can exceed $P^{\rm B}_{\rm Q} = 0.1$ only when
they are very close to the quasar model track or have near-IR magnitudes that exclude the case of Galactic dwarfs.
The HSC colors of the spectroscopically-confirmed galaxies are indistinguishable from those of the quasars, 
but most of the galaxies that we select are at $z \le 6$. 
This is presumably related to the faintness of these galaxies, with
magnitudes mostly around $z_{\rm AB} = 24.0$, i.e., close to the limit of our follow-up spectroscopy.
They would quickly become fainter if redshifted to $z > 6$,
as the observed $z$-band flux is progressively dominated by the GP trough.

The present quasar candidates also include an [\ion{O}{3}] emitter at $z = 0.954$, as displayed in Figure \ref{fig:spectra5}.
The strong [\ion{O}{3}] $\lambda$4959 and $\lambda$5007 lines in the $y$-band give this galaxy a red $z - y$ color. 
Table \ref{tab:spectroscopy} reports the measurements of the two [\ion{O}{3}] lines, H$\gamma$, and H$\beta$. 
Our past selection included many similar objects at $z \sim 0.8$, which appear as $i$-dropouts due to the strong lines in the $z$-band.
We now know that their colors are distinct from those of high-$z$ quasars, as evident in Figure \ref{fig:twocolor}, and thus we
recently incorporated an additional color cut ($i - z < 2.5$ and $z - y < -1.0$, as mentioned in \S \ref{sec:obs}) which eliminates most such contaminants.

We have also identified 31 Galactic cool stars and brown dwarfs with the present spectroscopy, as displayed in Figures \ref{fig:spectra6} and \ref{fig:spectra7}.
Table \ref{tab:bdtypes} lists the rough spectral classes, estimated by fitting the spectral standard templates of M4- to T8-type dwarfs 
\citep{burgasser14, skrzypek15} to the observed spectra. 
We emphasize that the stellar classifications are meant to be only approximate, given the relatively poor data quality and limited spectral coverage.
The number of Galactic dwarfs identified in SHELLQs has now grown to 96, many of which entered the sample because of 
inaccurate HSC photometry in past DRs. 
Indeed, the majority of the dwarfs have $P^{\rm B}_{\rm Q} < 0.1$ and thus would not be selected as candidates with the magnitudes from PDR3; see Figure \ref{fig:twocolor}.
The HSC-SSP data reduction pipeline \citep[{\it{hscPipe}};][]{bosch18} provides accurate flux measurements
for the vast majority of sources, but the high-$z$ quasars we are seeking are as rare as the very unusual cases of erroneous photometry that happened to dwarfs.
On a related note, the dwarfs we have identified spectroscopically represent a very biased sample, since the selection algorithm is tuned to remove typical dwarfs
with correct photometry.
Figure \ref{fig:twocolor} indicates that only the dwarfs closest to the quasar model track have $P^{\rm B}_{\rm Q} > 0.1$, 
as one would expect.\footnote{
The one exception is $J142331.18-010618.6$ with $i_{\rm AB} - z_{\rm AB} = 2.0$ and $z_{\rm AB} - y_{\rm AB} = 0.9$, which has
$P^{\rm B}_{\rm Q} = 1.0$ because of unexpectedly faint near-IR magnitudes given the HSC magnitudes.
}
Finally, given the limited data quality, we cannot exclude the possibility that some of these objects actually belong to extragalactic populations, e.g., compact and quiescent galaxies
at $z \sim 1$ which have similar colors to Galactic dwarfs; a further investigation on this issue requires much deeper observations than presented here.

The above tables and figures include
16 objects 
from our first set of spectroscopy of the low-$P^{\rm B}_{\rm Q}$ sample, which are deliberately selected from
the sources with $P^{\rm B}_{\rm Q} < 0.1$ (see Step 3 of the selection flow in \S \ref{sec:obs}).
Together with the main spectroscopic program for the $P^{\rm B}_{\rm Q} > 0.1$ candidates, these additional observations 
have completed the identification of 
PDR3 point sources with ($z_{\rm AB} < 24.0$,
$\sigma_z < 0.155$, $i_{\rm AB} - z_{\rm AB} > 2.0$, $z_{\rm AB} - y_{\rm AB} < 0.5$, and $y$-band detection; $i$-dropouts) across the entire Wide layer,
and those with ($y_{\rm AB} < 24.0$, $\sigma_y < 0.155$, and $z_{\rm AB} - y_{\rm AB} > 2.0$; $z$-dropouts) in the Spring 
fields (RA = 8$^{\rm h}$ -- 17$^{\rm h}$) of the Wide layer.\footnote{
In other words, there are only 16 objects that satisfy these color cuts and have $P^{\rm B}_{\rm Q} < 0.1$ over the above fields.}
We found that the 16 targets include 
three quasars at $z = 6.55 - 6.63$ 
($J091906.33+051235.3$, $J144045.91+001912.9$, and $J145005.39-014438.9$) and 13 Galactic dwarfs.
The redshift $z = 6.6$ corresponds to the quasar color $z_{\rm AB} - y_{\rm AB} \sim 2.0$, a boundary between the $P^{\rm B}_{\rm Q} > 0.1$ and $< 0.1$
regions of the two-color diagram 
(Figure \ref{fig:twocolor}), so discovery of the three quasars 
is perfectly consistent with what one would expect.
We see no evidence of significant incompleteness in the Bayesian selection compared to the simple color cuts.
 

As previously mentioned, a significant fraction of the contaminants in our spectroscopic sample comes from inaccurate HSC photometry in earlier DRs. 
A number of factors have contributed to the continuous improvement of the photometry, including the additional exposures of the same fields and the updated algorithms of
{\it hscPipe} for better treatments of, e.g., sky subtraction, removal of scattered light and artifacts, photometric calibration, and object detection
\citep{bosch18, aihara19, aihara21}.
If we had started the SHELLQs project with PDR3 photometry, the success rate of follow-up spectroscopy would have been much higher.
Out of the 312 spectroscopically-identified objects plotted in Figure \ref{fig:twocolor}, 205 pass our latest selection with the PDR3 photometry.
Of the 205 objects, 154 are high-$z$ quasars (including 22 type-II candidates)\footnote{
Figure \ref{fig:twocolor} clarifies what happened to the remaining eight quasars; three are too blue ($i_{\rm AB} - z_{\rm AB} < 1.5$) to pass the color cuts in Step 1 of the selection, 
and five have $P_{\rm Q}^{\rm B} < 0.1$ because of their colors relatively close to those of Galactic dwarfs.
}
and 33 are galaxies, amounting to fractions of 
75 \% (64 \% if the type-II candidates are excluded) and 16 \%, respectively.
Thus 91 \% of the photometric candidates lie at $z > 5.6$.
The [\ion{O}{3}] emitters make a negligible contribution (3 objects, 1 \%), while the contamination rate of Galactic dwarfs is 7 \% (15 objects).
Thus our selection algorithm has very high efficiency, even though we are probing an unprecedentedly low-luminosity population of quasars.

We are approaching the end of the HSC-SSP survey; as of Nov 2021, the survey has observed for 325 of the allocated 330 nights.
PDR3 includes the reduced data from 278 nights, and the spectroscopic identification of all but the faintest high-$z$ quasar candidates from the data 
have been reported in 
the previous and present SHELLQs papers.
The remaining candidates will be covered in our forthcoming observations, and eventually,
an SSP survey planned with the Subaru Prime Focus Spectrograph (PFS) under development \citep{takada14, tamura16, tamura18}
will provide the opportunity to observe a broader range of candidates, 
e.g., very faint ($z_{\rm AB} > 24.5$) or extended $i$-dropouts.
We aim to start the PFS-SSP survey in 2023.

The new discoveries reported here represent a significant increase in the size of the SHELLQs sample.
Our next goals are to tighten the constraints on the luminosity function (LF) at $z = 6$ \citep{p5} and measure the LF at $z = 7$.
Diverse follow-up projects are also ongoing, including near-zone size measurements \citep[e.g.,][T.-Y. Lu et al., in prep.]{ishimoto20}, clustering analyses, 
and characterization of the optical discovery spectra (A. Takahashi et al., in prep.) and near-IR broadband SEDs \citep[e.g.,][]{kato20},
as well as the two key projects with near-IR spectrographs and ALMA described in \S \ref{sec:intro}.
Further ambitious programs have been approved and await observations, including those with {\it Chandra} and {\it James Webb Space Telescope} \citep{onoue_jwst}
to study SHELLQs objects in greater detail.

\acknowledgments

This research is based on data collected at the Subaru Telescope, which is operated by the National Astronomical Observatory of Japan. 
We are honored and grateful for the opportunity of observing the Universe from Maunakea, which has the cultural, historical and natural significance in Hawaii.
We appreciate the staff members of the telescope for their support during our FOCAS observations.
The data analysis was in part carried out on the open use data analysis computer system at the Astronomy Data Center of NAOJ.

This work is also based on observations made with the GTC, installed at the Spanish Observatorio del Roque de los Muchachos 
of the Instituto de Astrof\'{i}sica de Canarias, on the island of La Palma.
We thank Stefan Geier and other support astronomers for their help during preparation and execution of our observing program.

Y. M. was supported by the Japan Society for the Promotion of Science (JSPS) KAKENHI Grant No. JP17H04830, No. 21H04494, and the Mitsubishi Foundation grant No. 30140.
K. I. acknowledges support by the Spanish MCIN under grant PID2019-105510GB-C33/AEI/10.13039/501100011033 and ``Unit of excellence Mar\'ia de Maeztu 2020-2023'' awarded to ICCUB (CEX2019-000918-M).

The HSC collaboration includes the astronomical communities of Japan and Taiwan, and Princeton University.  The HSC instrumentation and software were developed by the National Astronomical Observatory of Japan (NAOJ), the Kavli Institute for the Physics and Mathematics of the Universe (Kavli IPMU), the University of Tokyo, the High Energy Accelerator Research Organization (KEK), the Academia Sinica Institute for Astronomy and Astrophysics in Taiwan (ASIAA), and Princeton University.  Funding was contributed by the FIRST program from the Japanese Cabinet Office, the Ministry of Education, Culture, Sports, Science and Technology (MEXT), the Japan Society for the Promotion of Science (JSPS), Japan Science and Technology Agency  (JST), the Toray Science  Foundation, NAOJ, Kavli IPMU, KEK, ASIAA, and Princeton University.
 
This paper is based on data retrieved from the HSC data archive system, which is operated by Subaru Telescope and Astronomy Data Center (ADC) at NAOJ. 
Data analysis was in part carried out with the cooperation of Center for Computational Astrophysics (CfCA) at NAOJ.   

This paper makes use of software developed for Vera C. Rubin Observatory. We thank the Rubin Observatory for making their code available as free software at http://pipelines.lsst.io/. 
 
The Pan-STARRS1 Surveys (PS1) and the PS1 public science archive have been made possible through contributions by the Institute for Astronomy, the University of Hawaii, the Pan-STARRS Project Office, the Max Planck Society and its participating institutes, the Max Planck Institute for Astronomy, Heidelberg, and the Max Planck Institute for Extraterrestrial Physics, Garching, The Johns Hopkins University, Durham University, the University of Edinburgh, the Queen's University Belfast, the Harvard-Smithsonian Center for Astrophysics, the Las Cumbres Observatory Global Telescope Network Incorporated, the National Central University of Taiwan, the Space Telescope Science Institute, the National Aeronautics and Space Administration under grant No. NNX08AR22G issued through the Planetary Science Division of the NASA Science Mission Directorate, the National Science Foundation grant No. AST-1238877, the University of Maryland, Eotvos Lorand University (ELTE), the Los Alamos National Laboratory, and the Gordon and Betty Moore Foundation.


\begin{figure*}
\epsscale{1.0}
\plotone{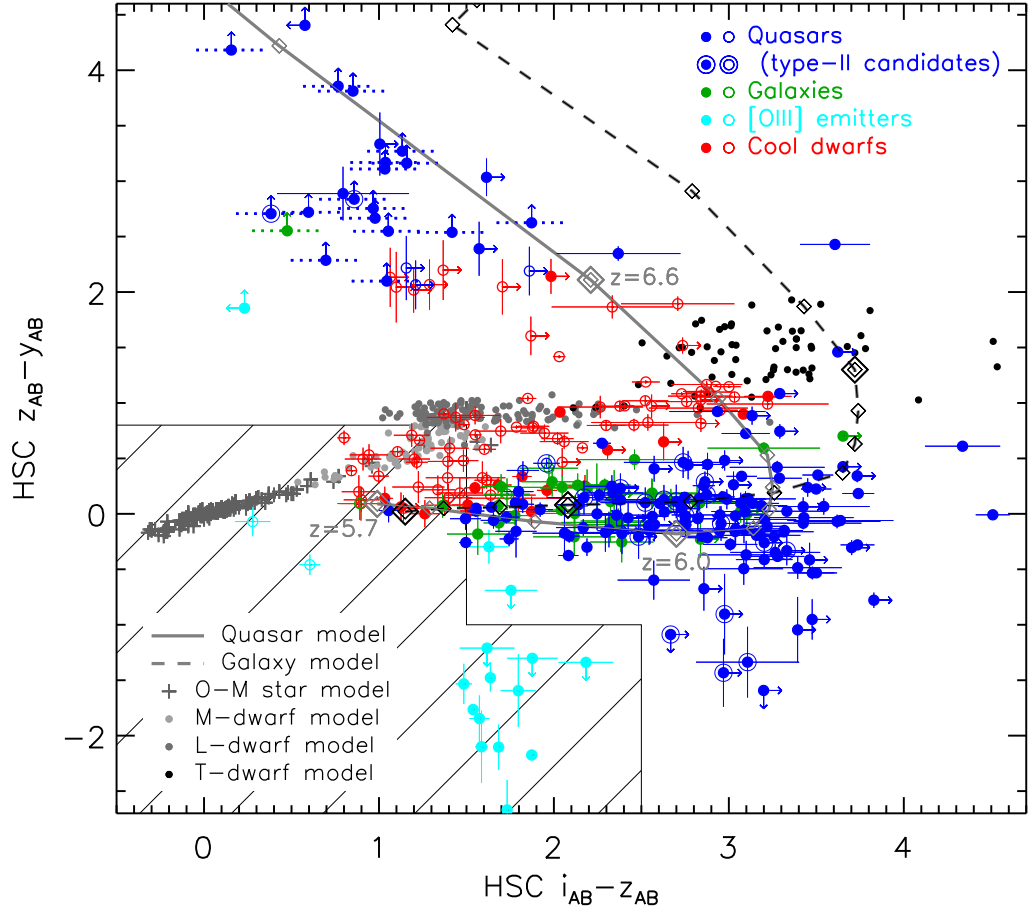}
\caption{HSC $i - z$ and $z - y$ colors of the high-$z$ quasars ({\it blue symbols}), the high-$z$ galaxies ({\it green}), 
the [\ion{O}{3}] emitters at $z = 0.8 - 1.0$ ({\it cyan}), and Galactic cool stars and brown dwarfs ({\it red}),
with spectroscopic identification reported in the past and present SHELLQs papers;
the filled and open circles represent those with $P_{\rm Q}^{\rm B} > 0.1$ and $P_{\rm Q}^{\rm B} \le 0.1$, respectively.
The magnitudes are taken from PDR3.
The arrows represent 3$\sigma$ limits, while the objects detected only in the $y$-band are placed at $i_{\rm AB} - z_{\rm AB} = i_{\rm AB}^{\rm lim} - z_{\rm AB}^{\rm lim}$ 
with dotted error bars, where $m_{\rm AB}^{\rm lim}$ represents the 3$\sigma$ limiting magnitude.
We also plot the model colors of quasars ({\it solid curve}) and galaxies ({\it dashed curve}) at $z \ge 5.7$, with the diamonds marking 
redshifts at $\Delta z = 0.1$ intervals.
The gray crosses and dots represent simulated colors of main-sequence stars and M-, L-, T-type brown dwarfs, as indicated at the bottom-left corner
\citep[see][for the details of those models]{p1}.
The hatched area represents the color cuts used in Step 1 of the selection (see \S \ref{sec:obs}).
\label{fig:twocolor}}
\end{figure*}

\clearpage

\startlongtable
\begin{deluxetable*}{rcccrrl}
\tablecaption{Journal of Discovery Spectroscopy \label{tab:obsjournal}}
\tablehead{
\colhead{Object} & 
\colhead{$i_{\rm AB}$} & \colhead{$z_{\rm AB}$} & \colhead{$y_{\rm AB}$} & \colhead{$P_{\rm Q}^{\rm B}$} & 
\colhead{$t_{\rm exp}$} & \colhead{Date (Inst)}\\
\colhead{} & 
\colhead{(mag)} & \colhead{(mag)} & \colhead{(mag)} & \colhead{} &
\colhead{(min)} & \colhead{}
} 
\startdata
\multicolumn{7}{c}{Quasars}\\\hline
$J131746.92+001722.2$ &          $>$26.02 &          $>$24.77 & 22.22 $\pm$ 0.03 &         1.000 &  120 & 2020 Apr 25 \& 30 (O)\\
$J134905.63+015608.9$ &          $>$26.00 &          $>$24.27 & 22.20 $\pm$ 0.07 &         1.000  & 120  &  2020 Mar 4 \& Apr 25 (O)\\
$J021430.90+023240.4$ & 26.41 $\pm$ 0.44 &          $>$25.49 & 21.86 $\pm$ 0.02 &         1.000 & 120  & 2020 Sep 17 -- 18 (O)\\
$J131050.13-005054.1$ & 27.06 $\pm$ 0.48 &          $>$25.16 & 22.55 $\pm$ 0.04 &         1.000  &  35 & 2021 Feb 21 (F)\\
$J021847.04+000715.0$ & 26.08 $\pm$ 0.23 &          $>$24.82 & 21.10 $\pm$ 0.01 &         1.000 &  50 &  2019 Oct 4 -- 6 (F)\\
$J133833.25-001832.9$ & 26.40 $\pm$ 0.29 & 25.61 $\pm$ 0.24 & 22.72 $\pm$ 0.04 &         1.000 & 30  & 2020 Dec 22 (F)\\
$J113034.65+045013.1$ & 27.71 $\pm$ 0.88 & 26.00 $\pm$ 0.68 & 22.78 $\pm$ 0.05 &         0.972 & 25  & 2021 Jan 5 (F)\\
$J225209.17+040243.8$ & 26.65 $\pm$ 0.54 & 25.89 $\pm$ 0.38 & 23.12 $\pm$ 0.08 &         0.403  & 30  & 2019 Oct 4 (F)\\
$J091041.14+005646.3$ & 26.78 $\pm$ 0.35 & 24.41 $\pm$ 0.06 & 22.06 $\pm$ 0.02 &         1.000 & 15  & 2020 Dec 22 (F)\\
$J103537.74+032435.7$ &          $>$25.88 & 24.99 $\pm$ 0.24 & 22.60 $\pm$ 0.05 &         0.111 & 30  & 2021 Jan 3 (F)\\
$J145005.39-014438.9$ &          $>$26.27 & 25.67 $\pm$ 0.28 & 23.45 $\pm$ 0.08 &         0.003  & 30  & 2021 Feb 21  (F)\\
$J102314.45-004447.9$ &          $>$26.10 & 25.11 $\pm$ 0.17 & 22.08 $\pm$ 0.02 &         1.000  &  15 & 2020 Dec 22 (F)\\
$J091906.33+051235.3$ & 26.83 $\pm$ 0.57 & 24.47 $\pm$ 0.21 & 22.28 $\pm$ 0.05 &         0.027  & 15  & 2021 Feb 22 (F)\\
$J111427.15+021517.6$ & 27.20 $\pm$ 0.44 & 23.36 $\pm$ 0.04 & 21.91 $\pm$ 0.02 &         1.000  & 15  & 2021 Jan 4 (F)\\
$J144045.91+001912.9$ & 27.12 $\pm$ 0.55 & 25.46 $\pm$ 0.21 & 23.39 $\pm$ 0.06 &         0.006  & 60  & 2021 Feb 23 \& Mar 3 (F)\\
$J110327.65+022947.2$ &          $>$25.97 & 23.39 $\pm$ 0.04 & 22.51 $\pm$ 0.07 &         0.910 & 15  & 2021 Jan 4 (F)\\
$J132539.95+441559.0$ & 25.42 $\pm$ 0.10 & 20.91 $\pm$ 0.01 & 20.92 $\pm$ 0.01 &     1.000 &  5 & 2021 Aug 16  (O)\\
$J132352.02-013941.4$ & 25.79 $\pm$ 0.27 & 22.85 $\pm$ 0.03 & 21.93 $\pm$ 0.03 &         1.000 & 30  & 2020 Mar 3 (O)\\
$J110423.75+024708.6$ & 27.13 $\pm$ 0.57 & 23.18 $\pm$ 0.03 & 22.94 $\pm$ 0.08 &         1.000 & 10  & 2021 Jan 4 (F)\\
$J140617.42+433033.2$ & $>$25.36              & 23.43 $\pm$ 0.04 & 23.46 $\pm$ 0.15 &      1.000  & 90  & 2021 Aug 12 (O)\\
$J105131.80-015159.0$ &          $>$26.00 & 23.79 $\pm$ 0.06 & 23.35 $\pm$ 0.11 &         1.000 &  30 & 2021 Jan 4 (F)\\
$J133206.46-011325.7$ & 24.94 $\pm$ 0.09 & 21.87 $\pm$ 0.01 & 21.53 $\pm$ 0.02 &         1.000 & 15  & 2020 Apr 30  (O)\\
$J093830.24+015405.8$ & 26.99 $\pm$ 0.42 & 23.82 $\pm$ 0.06 & 23.55 $\pm$ 0.10 &         1.000 & 30  & 2021 Jan 3 (F)\\
$J122331.91+025721.5$ & 26.00 $\pm$ 0.15 & 22.86 $\pm$ 0.03 & 22.77 $\pm$ 0.04 &         1.000  & 30  & 2020 Mar 3 (O)\\
$J012235.47-003602.4$ & 24.40 $\pm$ 0.07 & 21.14 $\pm$ 0.01 & 21.35 $\pm$ 0.01 &         1.000  &  15 & 2020 Sep 18  (O)\\
$J001650.69+034154.1$ & 27.04 $\pm$ 0.61& 23.33 $\pm$ 0.04 & 23.32 $\pm$ 0.06 &          1.000 & 90  & 2021 Aug 2 (O)\\
$J094713.88+034939.8$ & 27.12 $\pm$ 0.54 & 23.20 $\pm$ 0.04 & 24.15 $\pm$ 0.18 &         1.000 & 15  & 2021 Jan 3 (F)\\
$J104320.45+013104.7$ & 26.95 $\pm$ 0.50 & 22.94 $\pm$ 0.02 & 22.52 $\pm$ 0.04 &         1.000 &  25 & 2021 Feb 20 (F)\\
$J104311.09+011913.6$ & 26.25 $\pm$ 0.25 & 23.70 $\pm$ 0.03 & 23.85 $\pm$ 0.13 &         1.000  & 30  & 2021 Feb 20 (F)\\
$J132700.35+014147.7$ & 26.75 $\pm$ 0.53 & 23.50 $\pm$ 0.05 & 23.37 $\pm$ 0.08 &         1.000 & 35  & 2021 Jan 3 (F)\\
$J125845.66-004757.5$ & 27.32 $\pm$ 0.70 & 23.72 $\pm$ 0.05 & 23.55 $\pm$ 0.11 &         1.000   & 20  & 2021 Jan 2 (F)\\
$J084422.57+042353.7$ & 26.01 $\pm$ 0.25 & 22.61 $\pm$ 0.02 & 23.10 $\pm$ 0.10 &         1.000 & 15  & 2020 Dec 22 (F)\\
$J104429.18+010207.1$ & 26.91 $\pm$ 0.38 & 24.11 $\pm$ 0.04 & 24.13 $\pm$ 0.17 &         1.000 & 25  & 2021 Mar 3 (F)\\
$J104740.91+041724.3$ & 26.63 $\pm$ 0.26 & 23.09 $\pm$ 0.02 & 23.02 $\pm$ 0.07 &         1.000  & 15  & 2021 Jan 5  (F)\\
$J102047.40+042946.7$ & 25.15 $\pm$ 0.09 & 21.64 $\pm$ 0.01 & 21.29 $\pm$ 0.01 &         1.000  &  15 & 2020 Nov 18 (O)\\
$J093551.64+034847.2$ & 25.28 $\pm$ 0.08 & 22.24 $\pm$ 0.01 & 22.26 $\pm$ 0.03 &         1.000  & 15  & 2020 Nov 18 (O)\\
$J115924.01+020041.8$ & 27.11 $\pm$ 0.47 & 23.97 $\pm$ 0.06 & 24.64 $\pm$ 0.19 &         1.000  & 30  & 2021 Feb 20 (F)\\
$J011224.59-011557.3$ & 25.33 $\pm$ 0.27 & 22.38 $\pm$ 0.07 & 22.39 $\pm$ 0.04 &         1.000  &  90 & 2020 Dec 24 (F)\\
$J022223.74+024535.0$ & 24.72 $\pm$ 0.15 & 21.44 $\pm$ 0.01 & 21.82 $\pm$ 0.03 &         1.000  &  10 & 2020 Dec 22 (F)\\
$J112746.08+022231.5$ & 26.40 $\pm$ 0.22 & 23.41 $\pm$ 0.04 & 23.28 $\pm$ 0.08 &         1.000  & 30  & 2021 Jan 5 (F)\\
$J095139.01+021952.0$ &          $>$26.19 & 23.54 $\pm$ 0.05 & 25.79 $\pm$ 0.66 &         1.000  & 15  & 2021 Feb 21 (F)\\
$J001247.71+033716.9$ & 25.10 $\pm$ 0.10 & 21.37 $\pm$ 0.01 & 21.65 $\pm$ 0.01 &     1.000 & 15  & 2021 Aug 1 (O)\\
$J233429.70+050217.0$ & 24.68 $\pm$ 0.11 & 21.46 $\pm$ 0.02 & 21.57 $\pm$ 0.04 &      1.000  &  15 & 2021 Aug 1 (O)\\
$J002334.65+034223.2$ & 25.91 $\pm$ 0.19 & 22.82 $\pm$ 0.02 & 22.80 $\pm$ 0.04 &      1.000  &  30 & 2021 Aug 2 (O)\\
$J103734.52+003750.8$ & 26.38 $\pm$ 0.27 & 23.29 $\pm$ 0.04 & 23.66 $\pm$ 0.12 &         1.000  & 60  & 2020 Mar 3 (O)\\
$J102841.66+001755.9$ & 24.84 $\pm$ 0.07 & 21.87 $\pm$ 0.01 & 21.72 $\pm$ 0.02 &         1.000  & 15  & 2021 Feb 21 (F)\\
$J110248.67-010550.9$ & 25.05 $\pm$ 0.07 & 22.30 $\pm$ 0.01 & 22.37 $\pm$ 0.03 &         1.000  &  15 & 2021 Jan 24 (O)\\
$J111027.01+022800.7$ & 26.31 $\pm$ 0.25 & 22.69 $\pm$ 0.02 & 22.76 $\pm$ 0.06 &         1.000  & 15  & 2021 Jan 4 (F)\\
$J145520.26+031833.0$ & 25.75 $\pm$ 0.18 & 22.55 $\pm$ 0.02 & 22.60 $\pm$ 0.05 &         1.000  & 25  & 2021 Feb 20 (F)\\
$J110756.01-011819.0$ & 24.52 $\pm$ 0.04 & 21.77 $\pm$ 0.01 & 21.62 $\pm$ 0.02 &         1.000  & 15  & 2021 Jan 23 (O)\\
$J020415.74+001534.5$ & 25.42 $\pm$ 0.12 & 22.58 $\pm$ 0.02 & 22.78 $\pm$ 0.04 &         1.000 & 20  & 2020 Dec 22 (F)\\
$J132308.18+012619.2$ &          $>$25.45 & 23.80 $\pm$ 0.07 & 23.94 $\pm$ 0.14 &         1.000  & 20  &  2021 Jan 3 (F)\\
$J010718.87+002724.2$ & 25.12 $\pm$ 0.09 & 22.56 $\pm$ 0.03 & 22.58 $\pm$ 0.03 &         1.000  & 15  &  2019 Oct 4 (F)\\
$J120253.13+025630.8$ & 26.20 $\pm$ 0.17 & 23.66 $\pm$ 0.06 & 23.77 $\pm$ 0.10 &         1.000  & 15  & 2021 Feb 20 (F)\\
$J023551.42+013932.3$ & 24.46 $\pm$ 0.07 & 21.30 $\pm$ 0.01 & 21.48 $\pm$ 0.02 &         1.000 & 10  & 2020 Dec 22 (F)\\
$J110301.66+011845.7$ & 24.84 $\pm$ 0.07 & 22.92 $\pm$ 0.02 & 22.88 $\pm$ 0.06 &         1.000  & 15  & 2021 Feb 21 (F)\\
$J115006.96+021131.0$ & 24.52 $\pm$ 0.05 & 22.25 $\pm$ 0.02 & 21.61 $\pm$ 0.01 &         1.000  & 30  & 2021 Jan 2 (F)\\
$J110746.24+041101.2$ & 26.19 $\pm$ 0.20 & 24.02 $\pm$ 0.06 & 23.87 $\pm$ 0.14 &         0.994  &  25 & 2021 Mar 3 (F)\\
$J124359.03+032253.8$ & 26.24 $\pm$ 0.26 & 23.88 $\pm$ 0.06 & 23.89 $\pm$ 0.08 &         1.000  & 25  & 2021 Feb 21 (F)\\
$J010044.78+042211.5$ & 25.13 $\pm$ 0.15 & 22.96 $\pm$ 0.02 & 23.09 $\pm$ 0.07 &         1.000  &  45 & 2021 Aug 11 (O)\\
$J101505.59+000223.0$ & 24.04 $\pm$ 0.03 & 21.85 $\pm$ 0.01 & 22.15 $\pm$ 0.08 &         1.000  & 15  & 2021 Jan 24 (O)\\
$J091121.61+005146.1$ & 25.03 $\pm$ 0.07 & 23.30 $\pm$ 0.03 & 23.33 $\pm$ 0.05 &         1.000  & 25  & 2020 Dec 23 (F)\\
$J234836.96-003437.0$ & 23.88 $\pm$ 0.03 & 22.39 $\pm$ 0.02 & 22.65 $\pm$ 0.04 &         1.000   &  15 & 2019 Oct 5  (F)\\
$J145537.54+035929.0$ & 24.23 $\pm$ 0.06 & 22.15 $\pm$ 0.02 & 22.52 $\pm$ 0.05 &         1.000 & 25  & 2021 Feb 20 (F)\\
$J234901.28+005640.9$ & 24.40 $\pm$ 0.04 & 22.85 $\pm$ 0.02 & 22.82 $\pm$ 0.05 &         1.000  & 40  & 2020 Dec 22 (F)\\
$J131258.66+010902.1$ & 25.13 $\pm$ 0.12 & 22.98 $\pm$ 0.04 & 22.90 $\pm$ 0.09 &         1.000  &  15 & 2021 Jan 2 (F)\\
$J230610.54-012458.4$ & 24.62 $\pm$ 0.08 & 22.80 $\pm$ 0.04 & 22.41 $\pm$ 0.07 &         0.020  & 15  & 2019 Oct 5  (F)\\
$J102232.14+023041.9$ & 24.46 $\pm$ 0.06 & 22.64 $\pm$ 0.03 & 22.56 $\pm$ 0.07 &         1.000  & 15  & 2021 Jan 4 (F)\\
$J131732.73+012741.6$ & 23.70 $\pm$ 0.03 & 22.14 $\pm$ 0.02 & 22.09 $\pm$ 0.02 &         1.000  & 15  & 2020 May 16 (O)\\
\hline\multicolumn{7}{c}{Galaxies}\\\hline
$J132249.98+002849.4$ &          $>$25.89 &          $>$25.39 & 23.40 $\pm$ 0.08 &         1.000  &  50 & 2021 Feb 21 (F)\\
$J102936.55+040842.6$ & 26.36 $\pm$ 0.32 & 23.16 $\pm$ 0.04 & 22.57 $\pm$ 0.04 &         1.000 &  25 & 2021 Jan 3  (F)\\
$J102257.38-002608.2$ & 26.13 $\pm$ 0.25 & 24.01 $\pm$ 0.08 & 24.01 $\pm$ 0.14 &         0.912  &  25 & 2021 Mar 2 (F)\\
$J103553.04-010411.1$ & 26.69 $\pm$ 0.42 & 24.15 $\pm$ 0.07 & 24.30 $\pm$ 0.18 &         1.000  &  25 & 2021 Mar 3 (F)\\
$J095541.39+034025.0$ & 26.21 $\pm$ 0.23 & 24.09 $\pm$ 0.06 & 24.30 $\pm$ 0.16 &         1.000 & 25  & 2021 Mar 2 (F)\\
$J022443.33-064419.3$ & 24.96 $\pm$ 0.10 & 23.26 $\pm$ 0.05 & 23.02 $\pm$ 0.10 &         0.322 & 25  & 2019 Oct 5 (F)\\
$J021834.42-000426.1$ & 25.17 $\pm$ 0.10 & 23.81 $\pm$ 0.07 & 23.73 $\pm$ 0.09 &         0.381  & 30  & 2019 Oct 5 (F)\\
\hline\multicolumn{7}{c}{[\ion{O}{3}] Emitter}\\\hline
$J020816.31+000346.7$ & 25.68 $\pm$ 0.18 &          $>$24.89 & 23.59 $\pm$ 0.07 &         0.105 & 10  & 2019 Oct 5 (F)\\
\hline\multicolumn{7}{c}{Cool Dwarfs}\\\hline
$J000241.98+021705.6$ & 25.15 $\pm$ 0.08 & 22.97 $\pm$ 0.05 & 22.72 $\pm$ 0.09 &         1.000 & 45  & 2020 Sep 19 (O)\\
$J010231.32+044303.4$ & 25.29 $\pm$ 0.46 & 23.28 $\pm$ 0.05 & 23.21 $\pm$ 0.12 &        0.830  &  90 & 2021 Aug 3 (O)\\
$J010514.82+034839.9$ & $>$25.09              & 23.36 $\pm$ 0.05 & 22.79 $\pm$ 0.06 &         1.000  & 90  & 2021 Aug 12 (O)\\
$J015901.37+001043.0$ & 25.41 $\pm$ 0.10 & 24.01 $\pm$ 0.08 & 23.69 $\pm$ 0.07 &         0.000 & 60  & 2019 Oct 6  (F)\\
$J020446.06+021657.2$ & 24.51 $\pm$ 0.05 & 22.96 $\pm$ 0.03 & 22.72 $\pm$ 0.05 &         0.620  & 45  & 2020 Sep 20 (O)\\
$J021009.92-024422.0$ & 25.12 $\pm$ 0.04 & 23.16 $\pm$ 0.04 & 22.95 $\pm$ 0.05 &         1.000  & 25  & 2021 Jan 5 (F)\\
$J022812.09-012440.8$ & \nodata & 23.62 $\pm$ 0.04 & 23.24 $\pm$ 0.07 &         1.000  & 120  & 2020 Oct 10  (O)\\
$J023819.83-013259.7$ & 25.87 $\pm$ 0.29 & 24.07 $\pm$ 0.08 & 23.93 $\pm$ 0.18 &         0.019  & 70  & 2019 Oct 5 -- 6 (F)\\
$J083313.13+005505.5$ & 25.93 $\pm$ 0.38 & 23.82 $\pm$ 0.08 & 23.36 $\pm$ 0.10 &         0.000 & 75  & 2021 Feb 22 -- 23 (F)\\
$J092139.81+004932.0$ & 24.52 $\pm$ 0.04 & 22.67 $\pm$ 0.02 & 21.63 $\pm$ 0.01 &         0.000  & 15  & 2021 Feb 20 (F)\\
$J100500.97+003536.8$ & 27.07 $\pm$ 0.38 & 25.04 $\pm$ 0.15 & 22.90 $\pm$ 0.06 &         0.114 & 40  & 2020 Dec 22 (F)\\
$J113410.92+010951.1$ & 24.73 $\pm$ 0.05 & 23.21 $\pm$ 0.03 & 23.18 $\pm$ 0.06 &         1.000  & 25  & 2021 Feb 21 (F)\\
$J120457.75+005851.4$ &          $>$26.35 & 25.61 $\pm$ 0.21 & 23.55 $\pm$ 0.08 &         0.000 & 25  & 2021 Mar 2 (F)\\
$J120954.27-004105.7$ &          $>$26.45 & 25.91 $\pm$ 0.30 & 23.86 $\pm$ 0.10 &         0.000  & 30  & 2021 Mar 2 (F)\\
$J123559.91+031157.1$ &          $>$25.90 & 25.39 $\pm$ 0.26 & 23.26 $\pm$ 0.05 &         0.000 & 50  & 2021 Feb 22  (F)\\
$J124746.47+014812.9$ &          $>$25.99 & 25.35 $\pm$ 0.20 & 23.33 $\pm$ 0.06 &         0.000   & 25  & 2021 Feb 23 (F)\\
$J141128.46-014848.6$ & 27.07 $\pm$ 0.51 & 25.34 $\pm$ 0.26 & 23.14 $\pm$ 0.05 &         0.000  & 60  & 2021 Feb 22 (F)\\
$J142331.18-010618.6$ & 24.05 $\pm$ 0.03 & 22.01 $\pm$ 0.01 & 21.10 $\pm$ 0.01 &         1.000  & 15  & 2020 Apr 24 (O)\\
$J142536.16-010029.8$ & 25.05 $\pm$ 0.07 & 23.37 $\pm$ 0.04 & 23.11 $\pm$ 0.06 &         0.925  &  15 & 2021 Feb 21 (F)\\
$J162134.85+443104.1$ &          $>$25.83 & 24.77 $\pm$ 0.24 & 22.72 $\pm$ 0.07 &         0.001  &  15 & 2021 Feb 23 (F)\\
$J163105.93+432413.6$ & 26.11 $\pm$ 0.13 & 22.89 $\pm$ 0.02 & 21.83 $\pm$ 0.02 &         0.994  & 15  & 2021 Feb 20  (F)\\
$J220019.88+001100.9$ & 23.41 $\pm$ 0.03 & 21.25 $\pm$ 0.01 & 20.65 $\pm$ 0.02 &         0.000  & 15  &  2020 Aug 20 (O)\\
$J220142.44+023557.4$ & 24.48 $\pm$ 0.06 & 22.60 $\pm$ 0.03 & 21.82 $\pm$ 0.03 &         0.000  &  30 & 2020 Aug 20 (O)\\
$J220809.25+014202.4$ & 24.77 $\pm$ 0.05 & 22.95 $\pm$ 0.02 & 22.61 $\pm$ 0.03 &         1.000  & 45  & 2020 Sep 20 (O)\\
$J221417.79+010233.0$ & 25.43 $\pm$ 0.08 & 23.41 $\pm$ 0.04 & 22.73 $\pm$ 0.04 &         0.000  & 10  & 2019 Oct 5 (F)\\
$J223323.76+005633.7$ & 24.13 $\pm$ 0.02 & 22.34 $\pm$ 0.01 & 21.56 $\pm$ 0.01 &         0.000  & 15  & 2020 Aug 28 (O)\\
$J223718.63+052003.0$ & 25.82 $\pm$ 0.25 & 24.18 $\pm$ 0.08 & 24.01 $\pm$ 0.13 &         0.001  & 50  & 2019 Oct 6  (F)\\
$J224130.62+041820.2$ & 23.85 $\pm$ 0.03 & 22.36 $\pm$ 0.01 & 21.56 $\pm$ 0.02 &         0.000  & 15  & 2020 Aug 24 (O)\\
$J224538.06+010727.4$ & 24.16 $\pm$ 0.03 & 22.61 $\pm$ 0.02 & 21.89 $\pm$ 0.02 &         0.000  & 30  & 2020 Aug 20 (O)\\
$J224540.11+054300.0$ & 24.17 $\pm$ 0.06 & 22.57 $\pm$ 0.02 & 21.98 $\pm$ 0.03 &         0.000  & 23 & 2020 Aug 20 (O)\\
$J225149.04-011009.0$ & 25.29 $\pm$ 0.18 & 23.42 $\pm$ 0.08 & 23.40 $\pm$ 0.11 &         0.490  & 90  & 2020 Sep 18 (O)\\
\enddata
\tablecomments{
Coordinates are at J2000.0, and magnitude upper limits are placed at $5\sigma$ significance. 
We took magnitudes from the HSC-SSP PDR3, corrected for Galactic extinction,
and recalculated $P_{\rm Q}^{\rm B}$ for objects selected from earlier DRs. 
$J022812.09-012440.8$ has a significantly ($>5\sigma$) negative $i$-band flux record in the database, for unknown reasons (though it is clearly detected on the image), and thus $i_{\rm AB}$ is not reported.
The column ``$t_{\rm exp}$" reports the total exposure time for each obejct.
The instrument (Inst) ``F" and ``O" denote Subaru/FOCAS and GTC/OSIRIS, respectively.}
\end{deluxetable*}

\begin{deluxetable*}{ccccl}
\tablecaption{Objects detected in the near-IR bands \label{tab:nir_photometry}}
\tablehead{
\colhead{Name} & \colhead{$J_{\rm AB}$} & \colhead{$H_{\rm AB}$} & \colhead{$K_{\rm AB}$} & \colhead{Camera}\\
\colhead{} & \colhead{(mag)} & \colhead{(mag)} & \colhead{(mag)} & \colhead{} 
} 
\startdata
\multicolumn{5}{c}{Quasars}\\\hline
$J131050.13-005054.1$  &     \nodata           & 21.31 $\pm$ 0.22 &          \nodata      & VIRCAM\\
$J091041.14+005646.3$ & 22.01 $\pm$ 0.22 &          \nodata     & 21.92 $\pm$ 0.43   & VIRCAM \\
$J111427.15+021517.6$ & 21.92 $\pm$ 0.22 & 21.64 $\pm$ 0.28 & 21.17 $\pm$ 0.22   & VIRCAM \\
$J144045.91+001912.9$ &         \nodata       &         \nodata      & 21.72 $\pm$ 0.32  & VIRCAM\\
$J132352.02-013941.4$ & 22.36 $\pm$ 0.35 & 21.25 $\pm$ 0.20     &            \nodata      & VIRCAM\\
$J133206.46-011325.7$ & 21.08 $\pm$ 0.09 & 20.47 $\pm$ 0.16     & 19.88 $\pm$ 0.08  & VIRCAM\\
$J012235.47-003602.4$ & 21.34 $\pm$ 0.23  &            \nodata        &            \nodata     & VIRCAM\\
$J104320.45+013104.7$ & 21.95 $\pm$ 0.25 &            \nodata        & 20.99 $\pm$ 0.18  & VIRCAM\\
$J102047.40+042946.7$ & 21.04 $\pm$ 0.22 &            \nodata        &            \nodata     &  WFCAM\\
$J111027.01+022800.7$ & 22.25 $\pm$ 0.28 &            \nodata        &            \nodata      & VIRCAM\\
$J110756.01-011819.0$ & 21.94 $\pm$ 0.21 &            \nodata        &            \nodata       & VIRCAM\\
$J023551.42+013932.3$ &            \nodata      & 20.37 $\pm$ 0.23  & 19.99 $\pm$ 0.17   & WFCAM\\
$J115006.96+021131.0$ & 21.47 $\pm$ 0.16 & 20.68 $\pm$ 0.16   & 20.34 $\pm$ 0.11  & VIRCAM \\
$J131732.73+012741.6$ & 21.87 $\pm$ 0.25  &            \nodata      &            \nodata        & VIRCAM \\
\hline\multicolumn{5}{c}{Cool Dwarfs}\\\hline
$J092139.81+004932.0$ & 20.43 $\pm$ 0.07  & 19.27 $\pm$ 0.04  & 18.72 $\pm$ 0.03  & VIRCAM\\
$J113410.92+010951.1$ & 22.05 $\pm$ 0.22  &          \nodata      & 20.97 $\pm$ 0.18   & VIRCAM \\
$J120457.75+005851.4$ & 21.99 $\pm$ 0.24  &          \nodata      &          \nodata        & VIRCAM \\
$J123559.91+031157.1$ & 21.41 $\pm$ 0.18  &            \nodata      &             \nodata       & VIRCAM \\
$J124746.47+014812.9$ & 21.58 $\pm$ 0.16  &            \nodata       &            \nodata       & VIRCAM \\
$J141128.46-014848.6$ & 21.41 $\pm$ 0.14   &          \nodata       &          \nodata      & VIRCAM\\
$J142331.18-010618.6$ & 20.50 $\pm$ 0.08   &          \nodata       & 21.27 $\pm$ 0.25  & VIRCAM \\
$J220142.44+023557.4$ &            \nodata       & 19.86 $\pm$ 0.16   & 19.29 $\pm$ 0.08 & WFCAM\\
$J223323.76+005633.7$ & 20.40 $\pm$ 0.21  & 20.15 $\pm$ 0.17   & 19.97 $\pm$ 0.19  & WFCAM\\
$J224130.62+041820.2$ & 20.58 $\pm$ 0.26  &            \nodata        & 20.40 $\pm$ 0.26 & WFCAM \\
\enddata
\end{deluxetable*}

\clearpage

\startlongtable
\begin{deluxetable*}{ccccccc}
\tablecaption{Spectroscopic Properties\label{tab:spectroscopy}}
\tablehead{
\colhead{Name} & \colhead{Redshift} & 
\colhead{$M_{1450}$} & \colhead{Line} & 
\colhead{EW$_{\rm rest}$} & \colhead{FWHM} & 
\colhead{log $L_{\rm line}$}\\
\colhead{} & \colhead{} & 
\colhead{(mag)} & \colhead{} & 
\colhead{(\AA)} & \colhead{(km s$^{-1}$)} & 
\colhead{($L_{\rm line}$ in erg s$^{-1}$)}
} 
\startdata
\multicolumn{7}{c}{Quasars}\\\hline
$J131746.92+001722.2$ &  6.94 &   $-24.80 \pm 0.07$           & Ly$\alpha$ &     43 $\pm$ 3 &     1300 $\pm$ 300 &    44.45 $\pm$ 0.01 \\  
$J134905.63+015608.9$ &  6.94 &   $-24.83 \pm 0.06$           & Ly$\alpha$ &     20 $\pm$ 1 &     1100 $\pm$ 100 &    44.15 $\pm$ 0.01 \\  
$J021430.90+023240.4$ &  6.85 &   $-25.35 \pm 0.04$           & Ly$\alpha$ &      9 $\pm$ 2 &     6300 $\pm$ 2100 &    44.00 $\pm$ 0.08 \\   
$J131050.13-005054.1$ &  6.82 &   $-24.67 \pm 0.10$           & Ly$\alpha$ &     13 $\pm$ 3 &     4100 $\pm$ 1900 &    43.89 $\pm$ 0.08 \\  
                                       &       &                                                & \ion{N}{5} &      4 $\pm$ 1 &         990 $\pm$ 470 &    43.42 $\pm$ 0.08 \\  
$J021847.04+000715.0$ &  6.78 &   $-25.73 \pm 0.02$           & Ly$\alpha$ &     27 $\pm$ 2 &   16000 $\pm$ 5000 &    44.60 $\pm$ 0.03 \\  
$J133833.25-001832.9$ &  6.70 &   $-23.99 \pm 0.07$           & Ly$\alpha$ &      24 $\pm$ 3 &      1300 $\pm$ 500 &    43.87 $\pm$ 0.04 \\  
$J113034.65+045013.1$ &  6.68 &   $-23.94 \pm 0.10$           & Ly$\alpha$ &     25 $\pm$ 4 &     2100 $\pm$ 1400 &    43.86 $\pm$ 0.07 \\ 
$J225209.17+040243.8$ &  6.67 &   $-24.02 \pm 0.09$           & Ly$\alpha$ &     12 $\pm$ 1 &             $<$ 230        &    43.48 $\pm$ 0.03 \\  
$J091041.14+005646.3$ &  6.65 &   $-24.55 \pm 0.05$           & Ly$\alpha$ &     31 $\pm$ 2 &     1300 $\pm$ 1400 &    44.21 $\pm$ 0.03 \\ 
$J103537.74+032435.7$ &  6.63 &   $-24.51 \pm 0.08$           & Ly$\alpha$ &     21 $\pm$ 4 &       7400 $\pm$ 800 &    44.02 $\pm$ 0.08 \\  
$J145005.39-014438.9$ &  6.63 &   $-23.73 \pm 0.09$           & Ly$\alpha$ &     12 $\pm$ 1 &          900 $\pm$ 300 &    43.48 $\pm$ 0.04 \\  
$J102314.45-004447.9$ &  6.63 &   $-24.47 \pm 0.08$           & Ly$\alpha$ &     33 $\pm$ 5 &      4000 $\pm$ 2200 &    44.19 $\pm$ 0.06 \\  
$J091906.33+051235.3$ &  6.62 &   $-23.98 \pm 0.19$           & Ly$\alpha$ &   46 $\pm$ 12 &     6800 $\pm$ 1900 &    44.14 $\pm$ 0.08 \\  
$J111427.15+021517.6$ &  6.55 &   $-24.75 \pm 0.08$           & Ly$\alpha$ &     35 $\pm$ 4 &      9000 $\pm$ 300 &    44.32 $\pm$ 0.05 \\ 
$J144045.91+001912.9$ &  6.55 &   $-23.23 \pm 0.13$           & Ly$\alpha$ &       7 $\pm$ 2 &      1600 $\pm$ 600 &    43.05 $\pm$ 0.08 \\  
$J110327.65+022947.2$ &  6.49 &   $-24.18 \pm 0.07$           & Ly$\alpha$ &     52 $\pm$ 6 &     5400 $\pm$ 7100 &    44.27 $\pm$ 0.04 \\ 
$J132539.95+441559.0$ &  6.45 &   $-25.95 \pm 0.17$           & Ly$\alpha$ &     78 $\pm$ 14 &   7500 $\pm$ 4100 &    45.17 $\pm$ 0.03 \\  
$J132352.02-013941.4$ &  6.44 &    $-24.62 \pm 0.04$           & Ly$\alpha$ &     36 $\pm$ 2 &     6800 $\pm$ 1800 &    44.29 $\pm$ 0.02 \\ 
$J110423.75+024708.6$ &  6.41 &   $-24.06 \pm 0.08$           & Ly$\alpha$ &     41 $\pm$ 4 &      1800 $\pm$ 800 &    44.12 $\pm$ 0.04 \\ 
$J140617.42+433033.2$ &  6.39 &   $-23.75 \pm 0.17$           & Ly$\alpha$ &     48 $\pm$ 11 &     9900 $\pm$ 100 &    44.07 $\pm$ 0.07 \\ 
$J105131.80-015159.0$ &  6.34 &   $-23.41 \pm 0.11$           & Ly$\alpha$ &     24 $\pm$ 4 &      2800 $\pm$ 2200 &    43.62 $\pm$ 0.06 \\ 
$J133206.46-011325.7$ &  6.31 &   $-25.03 \pm 0.07$           & Ly$\alpha$ &      48 $\pm$ 5 &     6000 $\pm$ 5000 &    44.57 $\pm$ 0.03 \\  
$J093830.24+015405.8$ &  6.31 &   $-22.91 \pm 0.15$           & Ly$\alpha$ &    70 $\pm$ 12 &        300 $\pm$ 360 &    43.90 $\pm$ 0.05 \\ 
$J122331.91+025721.5$ &  6.27 &   $-24.26 \pm 0.26$              & \nodata &               \nodata &               \nodata &               \nodata \\ 
$J012235.47-003602.4$ &  6.26 &   $-25.08 \pm 0.09$           & Ly$\alpha$ &   120 $\pm$ 10 &     2000 $\pm$ 1600 &    45.00 $\pm$ 0.02 \\  
$J001650.69+034154.1$ &  6.26 &   $-23.64 \pm 0.09$           & Ly$\alpha$ &    33 $\pm$ 5 &       4600 $\pm$ 300 &     43.86 $\pm$ 0.06 \\ 
$J094713.88+034939.8$ &  6.25 &   $-22.77 \pm 0.14$           & Ly$\alpha$ &  150 $\pm$ 20 &        750 $\pm$ 350 &    44.19 $\pm$ 0.01 \\ 
$J104320.45+013104.7$ &  6.25 &   $-24.19 \pm 0.07$              & \nodata &             \nodata &               \nodata &               \nodata \\ 
$J104311.09+011913.6$ &  6.23 &   $-22.73 \pm 0.27$           & Ly$\alpha$ &    81 $\pm$ 22 &     3200 $\pm$ 1000 &    43.90 $\pm$ 0.04 \\ 
$J132700.35+014147.7$ &  6.23 &   $-23.36 \pm 0.15$           & Ly$\alpha$ &     33 $\pm$ 8 &       5100 $\pm$ 800 &    43.76 $\pm$ 0.08 \\ 
$J125845.66-004757.5$ &  6.22 &   $-22.99 \pm 0.11$           & Ly$\alpha$ &      34 $\pm$ 7 &         220 $\pm$ 150 &    43.70 $\pm$ 0.03 \\ 
$J084422.57+042353.7$ &  6.21 &   $-23.55 \pm 0.05$           & Ly$\alpha$ &   105 $\pm$ 5 &           590 $\pm$ 70 &    44.34 $\pm$ 0.01 \\ 
$J104429.18+010207.1$ &  6.20 &   $-23.29 \pm 0.07$           & Ly$\alpha$ &    11 $\pm$ 2 &           370 $\pm$ 420 &    43.16 $\pm$ 0.08 \\ 
$J104740.91+041724.3$ &  6.19 &   $-23.43 \pm 0.07$           & Ly$\alpha$ &    52 $\pm$ 6 &       7900 $\pm$ 2800 &    43.97 $\pm$ 0.04 \\ 
$J102047.40+042946.7$ &  6.18 &   $-25.36 \pm 0.02$           & Ly$\alpha$ &     14 $\pm$ 1 &        6400 $\pm$ 900 &    44.18 $\pm$ 0.04 \\  
$J093551.64+034847.2$ &  6.16 &   $-24.53 \pm 0.03$           & Ly$\alpha$ &     24 $\pm$ 2 &        4600 $\pm$ 600 &    44.08 $\pm$ 0.03 \\  
$J115924.01+020041.8$ &  6.15 &   $-22.31 \pm 0.12$           & Ly$\alpha$ &     57 $\pm$ 7 &          810 $\pm$ 460 &    43.58 $\pm$ 0.02 \\ 
$J011224.59-011557.3$ &  6.14 &   $-23.83 \pm 0.05$           & Ly$\alpha$ &     98 $\pm$ 7 &     12600 $\pm$ 7700 &    44.39 $\pm$ 0.02 \\  
$J022223.74+024535.0$ &  6.14 &   $-24.91 \pm 0.04$           & Ly$\alpha$ &     78 $\pm$ 4 &     1100 $\pm$ 3800 &    44.74 $\pm$ 0.01 \\  
$J112746.08+022231.5$ &  6.13 &   $-23.22 \pm 0.09$           & Ly$\alpha$ &     21 $\pm$ 3 &       4200 $\pm$ 400 &    43.52 $\pm$ 0.06 \\ 
$J095139.01+021952.0$ &  6.12 &   $-21.53 \pm 0.77$           & Ly$\alpha$ & 340 $\pm$ 240 &     1400 $\pm$ 700 &    44.04 $\pm$ 0.02 \\ 
$J001247.71+033716.9$ &  6.12 &   $-25.14 \pm 0.06$           & Ly$\alpha$ &     97 $\pm$ 33 &     1300 $\pm$ 500 &    44.72 $\pm$ 0.03 \\  
$J233429.70+050217.0$ &  6.12 &   $-25.15 \pm 0.02$           & Ly$\alpha$ &     38 $\pm$ 2 &     3100 $\pm$ 500 &     44.53 $\pm$ 0.02 \\  
$J002334.65+034223.2$ &  6.12 &   $-24.00 \pm 0.04$           & Ly$\alpha$ &     29 $\pm$ 3 &     7300 $\pm$ 300 &     43.96 $\pm$ 0.04 \\ 
$J103734.52+003750.8$ &  6.11 &   $-22.93 \pm 0.06$           & Ly$\alpha$ &     97 $\pm$ 6 &         820 $\pm$ 250 &    44.05 $\pm$ 0.01 \\ 
$J102841.66+001755.9$ &  6.10 &   $-24.94 \pm 0.02$           & Ly$\alpha$ &     14 $\pm$ 2 &   14100 $\pm$ 1700 &    44.00 $\pm$ 0.05 \\ 
$J110248.67-010550.9$ &  6.07 &   $-24.00 \pm 0.05$           & Ly$\alpha$ &     72 $\pm$ 5 &      4900 $\pm$ 2500 &    44.34 $\pm$ 0.02 \\  
$J111027.01+022800.7$ &  6.07 &   $-23.56 \pm 0.09$           & Ly$\alpha$ &     84 $\pm$ 9 &   11000 $\pm$ 2900 &    44.23 $\pm$ 0.03 \\ 
$J145520.26+031833.0$ &  6.06 &   $-23.82 \pm 0.05$           & Ly$\alpha$ &     43 $\pm$ 3 &     5600 $\pm$ 1100 &    44.04 $\pm$ 0.03 \\ 
$J110756.01-011819.0$ &  6.06 &   $-25.11 \pm 0.08$              & \nodata &               \nodata &               \nodata &               \nodata \\  
$J020415.74+001534.5$ &  6.05 &   $-24.16 \pm 0.03$           & Ly$\alpha$ &     16 $\pm$ 2 &          610 $\pm$ 80 &    43.76 $\pm$ 0.05 \\ 
$J132308.18+012619.2$ &  6.03 &   $-22.96 \pm 0.09$           & Ly$\alpha$ &     7 $\pm$ 1 &        1400 $\pm$ 200  &    42.95 $\pm$ 0.08 \\ 
$J010718.87+002724.2$ &  6.03 &   $-23.99 \pm 0.03$           & Ly$\alpha$ &     28 $\pm$ 1 &       5100 $\pm$ 900 &    43.94 $\pm$ 0.02 \\  
$J120253.13+025630.8$ &  6.02 &   $-22.78 \pm 0.14$           & Ly$\alpha$ &     28 $\pm$ 4 &         240 $\pm$ 180 &    43.56 $\pm$ 0.03 \\ 
$J023551.42+013932.3$ &  6.02 &   $-25.58 \pm 0.02$              & \nodata &            \nodata &               \nodata &               \nodata \\  
$J110301.66+011845.7$ &  6.00 &   $-23.53 \pm 0.08$           & Ly$\alpha$ &   39 $\pm$ 12 &        1700 $\pm$ 600 &    43.74 $\pm$ 0.04 \\ 
$J115006.96+021131.0$ &  6.00 &   $-24.57 \pm 0.07$              & \nodata &             \nodata &               \nodata &               \nodata \\ 
$J110746.24+041101.2$ &  5.99 &   $-23.18 \pm 0.10$           & Ly$\alpha$ &   15 $\pm$ 2   &         400 $\pm$ 60   &    43.20 $\pm$ 0.04 \\ 
$J124359.03+032253.8$ &  5.97 &   $-22.25 \pm 0.20$           & Ly$\alpha$ &   77 $\pm$ 17 &       6300 $\pm$ 400 &    43.68 $\pm$ 0.05 \\ 
$J010044.78+042211.5$ &  5.96 &   $-23.73 \pm 0.05$              & \nodata &               \nodata &               \nodata &               \nodata \\ 
$J101505.59+000223.0$ &  5.95 &   $-24.13 \pm 0.05$           & Ly$\alpha$ &   126 $\pm$ 7 &    15000 $\pm$ 1200 &    44.62 $\pm$ 0.01 \\  
$J091121.61+005146.1$ &  5.94 &   $-23.17 \pm 0.07$           & Ly$\alpha$ &     50 $\pm$ 5 &     1460 $\pm$     50 &    44.02 $\pm$ 0.02 \\ 
$J234836.96-003437.0$ &  5.94 &   $-24.05 \pm 0.03$           & Ly$\alpha$ &     45 $\pm$ 2 &          830 $\pm$ 420 &    44.13 $\pm$ 0.01 \\ 
                                          &       &                                         & \ion{N}{5}    &     36 $\pm$ 2 &       5100 $\pm$ 2100 &    44.02 $\pm$ 0.02 \\ 
$J145537.54+035929.0$ &  5.93 &   $-24.40 \pm 0.05$           & Ly$\alpha$ &     47 $\pm$ 4 &      1900 $\pm$ 1400 &    44.32 $\pm$ 0.04 \\ 
$J234901.28+005640.9$ &  5.92 &   $-23.79 \pm 0.06$           & Ly$\alpha$ &     15 $\pm$ 3 &      8000 $\pm$ 1100 &    43.60 $\pm$ 0.07 \\ 
$J131258.66+010902.1$ &  5.90 &   $-23.64 \pm 0.09$           & Ly$\alpha$ &     25 $\pm$ 4 &        6900 $\pm$ 400 &    43.74 $\pm$ 0.07 \\ 
$J230610.54-012458.4$ &  5.89 &   $-23.92 \pm 0.02$              & \nodata &              \nodata &               \nodata &               \nodata \\ 
$J102232.14+023041.9$ &  5.89 &   $-24.09 \pm 0.04$           & Ly$\alpha$ &     15 $\pm$ 3 &    11000 $\pm$ 3500 &    43.68 $\pm$ 0.08 \\ 
$J131732.73+012741.6$ &  5.88 &   $-24.39 \pm 0.23$              & \nodata &               \nodata &               \nodata &               \nodata \\ 
\hline\multicolumn{7}{c}{Galaxies}\\\hline
$J132249.98+002849.4$ &  6.66 &   $-23.69 \pm 0.07$              & \nodata &               \nodata &               \nodata &               \nodata \\  
$J102936.55+040842.6$ &  6.21 &   $-24.01 \pm 0.07$              & \nodata &               \nodata &               \nodata &               \nodata \\ 
$J102257.38-002608.2$ &  6.21 &   $-23.15 \pm 0.07$              & \nodata &               \nodata &               \nodata &               \nodata \\ 
$J103553.04-010411.1$ &  6.09 &   $-22.52 \pm 0.10$              & \nodata &               \nodata &               \nodata &               \nodata \\ 
$J095541.39+034025.0$ &  5.91 &   $-23.05 \pm 0.09$              & \nodata &               \nodata &               \nodata &               \nodata \\ 
$J022443.33-064419.3$ &  5.69 &   $-23.32 \pm 0.05$           & Ly$\alpha$ &           $>$ 17 &         380 $\pm$ 20 &          42.99 $\pm$ 0.03 \\ 
$J021834.42-000426.1$ &  5.68 &   $-22.92 \pm 0.08$              & \nodata &               \nodata &               \nodata &               \nodata \\ 
\hline\multicolumn{7}{c}{[\ion{O}{3}] Emitters}\\\hline
$J020816.31+000346.7$ & 0.954 &               \nodata            & H$\gamma$ &     86 $\pm$ 14 &             $<$ 230 &       40.87 $\pm$ 0.03 \\  
                      &       &                                    \nodata             & H$\beta$    &     158 $\pm$ 35 &     220 $\pm$ 90 &    41.14 $\pm$ 0.07 \\  
                      &       &                                    \nodata & [OIII] $\lambda$4959 &   271 $\pm$ 45 &             $<$ 230 &     41.38 $\pm$ 0.03 \\  
                      &       &                                    \nodata & [OIII] $\lambda$5007 &  696 $\pm$ 107 &             $<$ 230 &    41.78 $\pm$ 0.02 \\  
\enddata
\tablecomments{
The redshifts have uncertainties $\Delta z \sim 0.01 - 0.1$, depending on the spectral features around Ly$\alpha$; see text.
``EW$_{\rm rest}$" represents the rest-frame equivalent width; 3$\sigma$ upper limits are reported for objects without detectable continuum.}
\end{deluxetable*}


\begin{deluxetable}{cc}
\tablecaption{Spectral classes of the Galactic dwarfs\label{tab:bdtypes}}
\tablehead{
\colhead{Name} & \colhead{Class}
} 
\startdata
$J000241.98+021705.6$ & M8\\
$J010231.32+044303.4$ & M7\\
$J010514.82+034839.9$ & M8\\
$J015901.37+001043.0$ & M8\\
$J020446.06+021657.2$ & L3\\
$J021009.92-024422.0$ & M5\\
$J022812.09-012440.8$ & M7\\
$J023819.83-013259.7$ & M5\\
$J083313.13+005505.5$ & L7\\
$J092139.81+004932.0$ & T2\\
$J100500.97+003536.8$ & T6\\
$J113410.92+010951.1$ & L2\\
$J120457.75+005851.4$ & T8\\
$J120954.27-004105.7$ &  T3\\
$J123559.91+031157.1$ & T8\\
$J124746.47+014812.9$ & T5\\
$J141128.46-014848.6$ & T2\\
$J142331.18-010618.6$ & L1\\
$J142536.16-010029.8$ & M5\\
$J162134.85+443104.1$ & T6\\
$J163105.93+432413.6$ & T1\\
$J220019.88+001100.9$ & L9\\
$J220142.44+023557.4$ & L9\\
$J220809.25+014202.4$ & M8\\
$J221417.79+010233.0$ & M8\\
$J223323.76+005633.7$ & L7\\
$J223718.63+052003.0$ & M4\\
$J224130.62+041820.2$ & L8\\
$J224538.06+010727.4$ & M8\\
$J224540.11+054300.0$ & L1\\
$J225149.04-011009.0$ & M5\\
\enddata
\tablecomments{These classifications are only approximate; see text.}
\end{deluxetable}

\end{document}